\documentclass[acmsmall]{acmart}

\usepackage{cleveref}
\usepackage{xspace}
\usepackage[normalem]{ulem}
\usepackage{graphicx}
\usepackage[table,dvipsnames]{xcolor}
\usepackage{glossaries}
\usepackage{tabularx}
\usepackage{makecell}
\usepackage{caption}
\usepackage{subcaption}
\usepackage{algorithm}
\usepackage{algpseudocode}
\usepackage{amsmath} 
\usepackage{framed}
\usepackage{stfloats}
\usepackage[export]{adjustbox}
\usepackage{rotating}
\usepackage{enumitem}
\usepackage{multirow} 
\usepackage{bigfoot}
\usepackage{siunitx}

\newcommand{\keepcomment}{1} 
\newcommand{\isrevision}{0} 

\usepackage{nicematrix}

\ifnum\keepcomment=1
    
    \newcommand{\todo}[1]{\noindent{\textcolor{red}{\textbf{TODO:} #1}}}
    \newcommand{\mgnote}[1]{\noindent{\textcolor{blue}{\textbf{MAX:} #1}}}
    \newcommand{\pmnote}[1]{\noindent{\textcolor{teal}{\textbf{PM:} #1}}}
    \newcommand{\gcnote}[1]{{\leavevmode\color{DarkOrchid}{\textbf{GAB:} #1}}}
    \newcommand{\mfnote}[1]{{\leavevmode\color{Sepia}{\textbf{MF:} #1}}}
    \newcommand{\rmnote}[1]{{\leavevmode\color{}{\textbf{RM:} #1}}}

    \newcommand{\delete}[1]{\leavevmode\sout{#1}}
    
    \newcommand{\gcdelete}[1]{\textcolor{DarkOrchid}{\leavevmode\sout{#1}}}
    
    \newcommand{\stkout}[1]{\ifmmode\text{\sout{\ensuremath{#1}}}\else\sout{#1}\fi}

    \newcommand{\sm}[1]{\textcolor{orange}{#1}}
    \newcommand{\pmed}[1]{\textcolor{red}{#1}}
\else
    \newcommand{\todo}[1]{\leavevmode\ignorespaces\unskip}
    \newcommand{\mgnote}[1]{\leavevmode\ignorespaces\unskip}
    \newcommand{\pmnote}[1]{\leavevmode\ignorespaces\unskip}
    \newcommand{\gcnote}[1]{\leavevmode\ignorespaces\unskip}
    \newcommand{\mfnote}[1]{\leavevmode\ignorespaces\unskip}
    \newcommand{\rmnote}[1]{\leavevmode\ignorespaces\unskip}
    
    \newcommand{\delete}[1]{\leavevmode\ignorespaces\unskip}
	\newcommand{\gcdelete}[1]{\leavevmode\ignorespaces\unskip}
	
    \newcommand{\sm}[1]{\leavevmode\ignorespaces\unskip}
    \newcommand{\pmed}[1]{\leavevmode\ignorespaces\unskip}
\fi

\ifnum\isrevision=1
    
\else
    
\fi

\newcommand{\eat}[1]{}
\newcommand{\eg}{{e.g.,}\xspace}
\newcommand{\ie}{{i.e.,}\xspace}

\newacronym{mccl}{MCCL}{Measurement-based Collective Communication Library}
\newacronym[
    longplural={Collective Communication Libraries},
    plural={CCLs}
]{ccl}{CCL}{Collective Communication Library}
\newacronym{nccl}{NCCL}{Nvidia CCL}
\newacronym{rccl}{RCCL}{AMD ROCm CCL}
\newacronym{hccl}{HCCL}{Huawei CCL}
\newacronym{smt}{SMT}{Satisfiability Modulo Theories}
\newacronym{milp}{MILP}{Mixed Integer Linear Program}
\newacronym{nic}{NIC}{Network Interface Card}
\newacronym{dl}{DL}{Deep Learning}
\newacronym{sm}{SM}{Streaming Multiprocessor}
\newacronym{tb}{TB}{Thread Block}
\newacronym{bfs}{BFS}{Breadth-First Search}
\newacronym{cg}{CG}{Column Generation}
\newacronym{hbm}{HBM}{High Bandwidth Memory}
\newacronym{lp}{LP}{Linear Program}
\newacronym{fsdp}{FSDP}{Fully Sharded Data Parallelism}
\newacronym{rmp}{RMP}{Restricted Master Problem}
\newacronym{ewma}{EWMA}{Exponentially Weighted Moving Average}
\glsdisablehyper

\DeclareSIUnit{\Kbps}{Kbps}
\DeclareSIUnit{\Mbps}{Mbps}
\DeclareSIUnit{\Gbps}{Gbps}

\allowdisplaybreaks

\newenvironment{smallalign}
  {%
    \begingroup
    \edef\savedbaselineskip{\the\baselineskip}%
    \footnotesize
    \setlength{\baselineskip}{\savedbaselineskip}%
    \align
  }
  {%
    \endalign
    \endgroup
  }

\newenvironment{smallmath}
  {%
    \begingroup
    \edef\savedbaselineskip{\the\baselineskip}%
    \footnotesize
    \setlength{\baselineskip}{\savedbaselineskip}%
    \[
  }
  {%
    \]
    \endgroup
  }

\begin{document}

\setcopyright{cc}
\setcctype{by}
\acmJournal{PACMNET}
\acmYear{2026} \acmVolume{4} \acmNumber{CoNEXT4} \acmArticle{55}
\acmMonth{12} \acmDOI{10.1145/3856969}

\received{5 June 2026}
\received[accepted]{10 September 2026}


\title[MCCL: Measurement-based Collective Communication Library]{MCCL: Measurement-based Collective Communication Library}


\author{Michele Ferrero}
\affiliation{%
  \institution{EURECOM}
  \city{Biot}
  \country{France}
}
\email{ferrerom@eurecom.fr}

\author{Gabriele Castellano}
\affiliation{%
  \institution{Independent Researcher}
  \city{Paris}
  \country{France}
}
\email{gcastellano.phd@gmail.com}

\author{Paolo Medagliani}
\affiliation{%
  \institution{Independent Researcher}
  \city{Paris}
  \country{France}
}
\email{paolomeda@gmail.com}

\author{Sebastien Martin}
\affiliation{%
  \institution{LCOMS, Universit{\'e} de Lorraine}
  \city{Metz}
  \country{France}
}
\email{martin13@univ-lorraine.fr}

\author{Roberto Morabito}
\affiliation{%
  \institution{EURECOM}
  \city{Biot}
  \country{France}
}
\email{roberto.morabito@eurecom.fr}

\author{Massimo Gallo}
\affiliation{%
  \institution{Independent Researcher}
  \city{Paris}
  \country{France}
}
\email{gallomassimo3@gmail.com}

\renewcommand{\shortauthors}{Michele Ferrero et al.}

\begin{abstract}
Collective communication is a critical performance bottleneck in large-scale distributed training. Existing libraries and synthesis frameworks compute \textit{open-loop} schedules once, assuming a static topology and peak link capacities. In production, however, persistent gray failures can silently degrade link throughput and latency for hours, causing pre-computed schedules to lose much of their efficiency during execution. In this paper, we present \gls{mccl}, a collective communication library that closes the loop between runtime network measurement and dynamic schedule synthesis. \gls{mccl} passively measures GPU-to-GPU throughput, detects persistent degradation, re-synthesizes the affected communication-intensive schedules with warm-started synthesizers, and hot-swaps them without interrupting training. We validate the complete loop on a multi-node hardware prototype and use simulations to evaluate cluster-scale scenarios. In our experiments, \gls{mccl} incurs negligible overhead ($\mathbf{{<}\,2\%}$) while maintaining high measurement accuracy ($\mathbf{{<}\,1.2\%}$ error) and low schedule hot-swap time (${\approx}\,\textbf{\SI{12}{\milli\second}}$). Notably, under gray failures, \gls{mccl} reduces training time by up to $\mathbf{4.99\times}$ over static schedules on simulated training workloads for state-of-the-art LLMs such as Llama and DeepSeek. 
\end{abstract}
\glsresetall

\maketitle

\section{Introduction}

Large-scale distributed training has become the foundation of modern deep learning, powering state-of-the-art models in natural language processing, computer vision, and recommendation systems. As model parameters and dataset scales continue to grow, the end-to-end efficiency of distributed training increasingly depends on the performance of collective communications. Primitives such as \textsc{AllGather}, \textsc{AllToAll}, \textsc{AllReduce}, and \textsc{ReduceScatter} directly influence training time, often accounting for a substantial fraction of the overall iteration latency. Consequently, significant research has focused on optimizing these primitives, with efforts largely concentrated on communication-computation overlap, reduction algorithms, and static topology-aware scheduling.

Hardware accelerator vendors provide \glspl{ccl} such as \gls{nccl} \cite{nccl}, \gls{rccl} \cite{rccl}, \gls{hccl} \cite{hccl}, and Intel OneCCL~\cite{oneccl}, with highly optimized collective communication schedules tailored to common cluster topologies and hardware configurations. These vendor libraries increasingly leverage specialized hardware, such as in-network computing~\cite{sharp}, to accelerate \textsc{Reduce}-intensive collectives. In parallel, schedule synthesizers \cite{shah2023taccl, xuting2024teccl, zhao2025forestcoll, jiamin2025syccl, zixian2021sccl, rashidi2022themis, wang2020blink} have demonstrated substantial speedups on communication-intensive primitives such as \textsc{AllGather} and \textsc{AllToAll}. However, existing synthesis tools share a critical limitation: they operate in an \textit{open-loop}, relying on static topology assumptions and peak link capacities that rarely hold in production environments.

Datacenter networks typically exhibit complex and dynamic behavior. A critical yet pervasive challenge is the prevalence of \textit{gray failures} ~\cite{huang2017gray,haryadi2018failslow}, where components remain functionally operational but experience severe performance degradation due to thermal throttling, PCIe or \gls{nic} symbol errors, and optical link decay. Unlike \textit{hard failures}, which typically trigger alarms, gray failures can silently reduce link performance, sometimes substantially, for several minutes or even hours \cite{haryadi2018failslow, wu2025greyhound}. Consequently, optimal collective communication schedules at the beginning of training can become increasingly inefficient as system conditions evolve, significantly slowing training processes and leading to suboptimal resource utilization across the distributed infrastructure. This mismatch is particularly pronounced in iterative training because the exact same collectives repeat for thousands of iterations; any localized performance degradation becomes persistent and cumulative. 

Our key observation is that the repetitive nature of distributed training turns every collective into a free measurement opportunity. Because identical GPU-to-GPU transfers repeat across thousands of iterations, their progress can be passively monitored at path granularity, without injecting any traffic. This insight enables a fundamentally different operating model for collective communication. Instead of synthesizing a schedule once under static topology assumptions and peak capacities (\emph{open-loop}), the library can continuously \emph{measure} the paths it uses, \emph{decide} when the deployed schedule diverges from actual network conditions, \emph{re-synthesize} it for observed capacities, and \emph{deploy} the new schedule dynamically during training (\emph{closed-loop}).

Realizing this closed-loop architecture introduces three fundamental challenges. First, the system requires path-level visibility without perturbing training. Existing monitoring approaches either rely on synthetic probes, which compete with training traffic, or on aggregate collective completion time, which can detect slowdown but cannot localize the degraded paths. Second, to avoid triggering spurious re-synthesis cycles, the system must reliably distinguish persistent degradations from short-lived fluctuations and pinpoint the affected collectives. Finally, re-synthesis time is the primary barrier to rapid runtime adaptation. Tree-packing synthesizers such as ForestColl~\cite{zhao2025forestcoll} can quickly generate \textsc{AllGather} and \textsc{ReduceScatter} schedules, yet they lack support for warm-starting; even after an isolated, single-link degradation, they discard structurally valid subtrees and recompute schedules from scratch. The barrier is even higher for \textsc{AllToAll}: existing synthesizers~\cite{xuting2024teccl} solve heavyweight \gls{lp} formulations that take hours to converge, too slow for online adaptation. While recent approaches accelerate search times by exploiting network symmetry~\cite{jiamin2025syccl, shapley2026optccl}, gray failures inherently break topological symmetry, invalidating their core performance assumptions.

In this paper, we present \acrfull{mccl}\anon[]{\footnote{MCCL is pronounced ``Mickel''.}}, a collective communication library that integrates a closed optimization loop directly into the communication stack. \gls{mccl} employs a controller-executor architecture: distributed executors run collectives and passively collect GPU-to-GPU telemetry on the fast path, while a centralized controller detects persistent performance drift, selectively re-synthesizes the affected collectives, and deploys the resulting schedules through a hot-swap mechanism adapted from MCCS~\cite{Yongji2024mccs}. Each new component of \gls{mccl} is designed around the requirements of the loop rather than as a standalone optimization. \gls{mccl} targets communication-intensive collectives, namely \textsc{AllGather} and \textsc{AllToAll}. Reduction-centric collectives such as \textsc{ReduceScatter} keep their existing highly optimized implementations, avoiding interference with in-network computing. \textsc{AllReduce} remains compatible with \gls{mccl} through its standard decomposition into \textsc{ReduceScatter} and \textsc{AllGather}, optimizing its \textsc{AllGather} phase while leaving the reduction phase unchanged.

In summary, this work makes the following contributions:

\begin{itemize}
    \item We design and implement \gls{mccl}, the first collective communication library that closes the loop between path-level measurements and schedule synthesis, continuously adapting communication-intensive schedules to live network conditions without interrupting ongoing training processes (\Cref{sec:system}).
    \item To fill the gap of existing synthesizers, we present a warm-started version of ForestColl that reuses the structure of deployed \textsc{AllGather} schedules, and a column-generation formulation for \textsc{AllToAll} 
    whose rounding gap stays within $\textbf{\SI{2.13}{\percent}}$ across tested topologies (\Cref{sec:optimization}). 
    \item To enable reliable path-level visibility, we design a passive telemetry framework that captures GPU-to-GPU bandwidth metrics at the thread-block level, with ${<}\,\textbf{\SI{2}{\percent}}$ overhead and ${<}\,\textbf{\SI{1.2}{\percent}}$ measurement error (\Cref{sec:measurements}).
    \item We validate the complete loop (telemetry, failure detection, re-optimization, and hot-swapping) on an 8-GPU, 2-node prototype, where a schedule swap completes in ${\approx}\,\textbf{\SI{12}{\milli\second}}$ without impacting training-related collective communication scheduling (\Cref{sec:results}). 
    \item We evaluate \gls{mccl} at cluster-scale simulating state-of-the-art LLM trainings, demonstrating up to $\mathbf{4.99\times}$ faster training under gray failures compared to static schedules (\Cref{sec:results}).
\end{itemize}

The remainder of this paper is organized as follows. We provide background in \Cref{sec:background}, 
\Cref{sec:system,sec:optimization,sec:measurements} 
detail the system architecture, optimization formulation, and measurement framework. Finally, we evaluate MCCL in \Cref{sec:results}, discuss related work in \Cref{sec:rw}, and conclude in \Cref{sec:conclusion}.

\section{Background and motivation}
\label{sec:background}



\subsection{Collective Communication Synthesis}


The collective communication research landscape~\cite{weingram2023xccl} is rapidly evolving, with hand-tuned vendor libraries increasingly complemented by automated schedule-synthesis frameworks. Libraries provided by hardware accelerator vendors, like \gls{nccl}~\cite{nccl}, \gls{rccl}~\cite{rccl}, \gls{hccl}~\cite{hccl}, and OneCCL~\cite{oneccl}, rely on static, heuristic algorithms (e.g., Ring, Tree) mapped to specific hardware topologies. While these libraries execute reduction-heavy collectives (e.g., \textsc{AllReduce}) with high efficiency by leveraging specialized hardware such as SHARP in-network computing~\cite{sharp}, the benefits of such accelerators are largely confined to collectives that involve reduction operations. Communication-intensive collectives, such as \textsc{AllGather} and \textsc{AllToAll}, remain primarily constrained by data movement and often cannot fully exploit heterogeneous or non-standard interconnects. Optimizing these collectives is therefore critical to reducing communication overhead and, ultimately, training time. To enable this flexibility, platforms such as MSCCL~\cite{msccl} were developed, enabling the execution of custom collective algorithms on top of standard libraries such as \gls{nccl}. 

\begin{figure}[t!]
    \begin{minipage}[b]{0.49\linewidth}
        \captionsetup*{type=table}
        \vspace{0cm}
        \tiny
        \centering
        \renewcommand{\arraystretch}{1.1}
        \begin{NiceTabular}{|w{c}{0.95cm}w{c}{0.7cm}w{c}{1.5cm}w{c}{1.7cm}|}[cell-space-limits=0.7pt]
            \CodeBefore
                \rowcolors{3}{gray!10}{white}
            \Body
            \hline
            \textbf{Name} & \textbf{Synthesis} & \textbf{Optimization} & \textbf{Re-optimization} \\
            \hline
            \Block{}{NCCL\cite{nccl}, \\ RCCL\cite{rccl}} & Static & N.A. & $\times$ \\
            \Block{}{SCCL\cite{zixian2021sccl}, \\ TACCL\cite{shah2023taccl}, \\ TECCL\cite{xuting2024teccl}} & Offline & \Block{}{Static, \\ peak topology} & Hard \\
            ForestColl\cite{zhao2025forestcoll} & Offline & \Block{}{Static, \\ peak topology} & \Block{}{\textsc{AllGather} \\ no \textsc{AllToAll}} \\
            SyCCL\cite{jiamin2025syccl} & Offline & \Block{}{Static, \\ peak topology} & \Block{}{Hard \\ with asymmetry} \\
            OptCCL\cite{shapley2026optccl} & Offline & \Block{}{Static, \\ peak topology} & \Block{}{Slow\\ with asymmetry} \\
            \hline
            MCCL & Online & \Block{}{Dynamic, \\ measured throughput} & Possible \\
            \hline
        \end{NiceTabular}
        \caption{Collective Communication Algorithms.}
        \label{tab:ccls-algo}
    \end{minipage}
    \hfill
    \begin{minipage}[b]{0.49\linewidth}
        \captionsetup{type=figure}
        \vspace{-2pt}
        \centering
        \includegraphics[width=\linewidth]{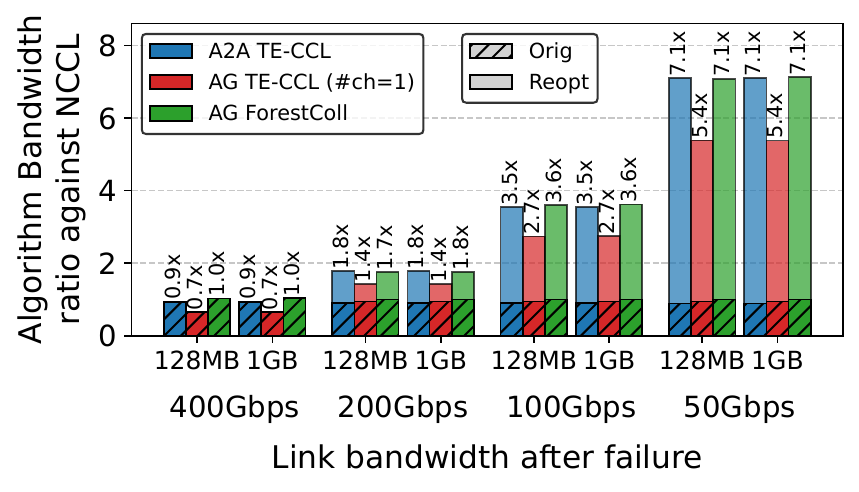}
        \caption{\gls{ccl} scheduling algorithm performances.} 
        \label{fig:motivation}
    \end{minipage}
\end{figure}

Building on this foundation, several collective communication synthesizers have emerged to automate schedule generation. Early approaches such as SCCL~\cite{zixian2021sccl} use \gls{smt} solvers to find Pareto-optimal algorithms, while TACCL~\cite{shah2023taccl} uses user-provided ``communication sketches'' to guide synthesis. Subsequent frameworks adopted traffic-engineering principles: TECCL~\cite{xuting2024teccl} formulates \textsc{AllGather} as an $A^*$-scaled \gls{milp} and \textsc{AllToAll} as an \gls{lp}, while ForestColl~\cite{zhao2025forestcoll} constructs spanning tree packs using strongly polynomial-time algorithms to generate throughput-optimal schedules. More recently, frameworks have focused on scaling these optimizations for massive topologies. Concurrent work OptCCL~\cite{shapley2026optccl} and SyCCL~\cite{jiamin2025syccl} both leverage network and collective symmetry to decompose the scheduling problem, with OptCCL explicitly separating routing from chunk scheduling via a multi-commodity flow \gls{lp}. However, a critical common denominator across these approaches (cf. \Cref{tab:ccls-algo}) is the ``static world'' assumption: they optimize for a topology snapshot taken at initialization time that is assumed to remain constant throughout training.



\subsection{Challenges: Scale and Reliability}
\label{subsec:challenges}

Training foundation models like LLama3 \cite{grattafiori2024llama3}, Gemini 2.5 \cite{comanici2025gemini25}, DeepSeek \cite{liu2024deepseek} involves big clusters with over 10k GPUs/TPUs running continuously for weeks or months. Furthermore, multiple major AI labs, including OpenAI, Microsoft, xAI, and Meta, are racing to deploy individual training clusters exceeding 100k accelerators, pushing datacenters scale to unprecedented levels. At this scale, network-level component failures (\eg links, \glspl{nic}, and optical transceivers), which might be considered rare exceptions in isolation, become a near certainty and pose a critical operational challenge~\cite{kun2024hpn}. For instance, estimates suggest that even if each \gls{nic}-to-switch link had a mean time to failure of 5 years, the first training job failure would likely occur in less than half an hour within a 100k GPU cluster~\cite{patel2024semianalysis}. Such high hardware failure rates are particularly problematic in distributed training, where a single component failure can cause the entire training job to halt, leaving expensive accelerators idle. 

While catastrophic fail-stop events are relatively easy to detect, a more challenging issue is the presence of \textit{gray failures} (or \textit{fail-slow} hardware) \cite{huang2017gray,haryadi2018failslow}, where components remain functional but operate at degraded performance levels due to specific recovery mechanisms \cite{leon2022jupiter,bonato2025reps,sobhani2025prime}. These performance degradations are rarely brief anomalies; empirical evidence shows that congestion-induced gray failures persist for an average of 72 minutes (e.g., Greyhound~\cite{wu2025greyhound}), while hardware-related faults can last significantly longer (\eg{} FailSlow~\cite{haryadi2018failslow}). Because these degraded states persist across thousands of training iterations, they become severe long-term bottlenecks.


\subsection{The case for dynamic CCL schedules}



Given these reliability challenges, we argue for a shift from static planning to dynamic adaptation, as pre-optimized schedules become highly inefficient when communication performance drifts. 

To motivate dynamic re-synthesis, we simulated \textsc{AllGather}, and \textsc{AllToAll} collectives under varying link-layer degradations using SimAI~\cite{wang2025simai, simai}, a high-fidelity simulator validated against production clusters (iteration-time and collective-communication deviations below $\SI{3.9}{\percent}$~\cite{wang2025simai}). We extend\footnote{\url{https://github.com/NetMeasurements-Team/SimAI}.} SimAI to support custom \gls{ccl} schedules using the same validated point-to-point primitives as its native \gls{nccl} algorithms, thereby preserving the simulator's reported accuracy, as confirmed by our custom ring schedule reproducing SimAI's native ring completion times.
We evaluate a 2-host (16 GPUs) rail-optimized topology (cf. \Cref{{app:topology}}, \Cref{fig:railtopology} for reference), using \gls{nccl}, TECCL, and ForestColl as representatives of state-of-the-art \gls{ccl}. Synthesizers are evaluated with up to ten chunks, reporting the best-performing configuration in each case. TECCL's \textsc{AllGather} is limited to one chunk because of scalability issues. 

As shown in \Cref{fig:motivation}, with peak capacity \ie{} all links connecting the nodes to the switch have \SI{400}{\Gbps}, TECCL and ForestColl match \gls{nccl} Ring's Algorithm Bandwidth~\cite{algobw}, which is already throughput optimal, with marginal deviations originating from chunking approximation errors. However, when simulating a gray failure by degrading the link capacity of GPU 0 by 50\%, 75\%, and 88\%, the performance of all static schedules (Orig) collapses to the speed of the bottleneck. Strikingly, \Cref{fig:motivation} demonstrates that adapting the schedule via re-optimization yields up to a $7.1\times$ improvement over these static baselines.

\subsection{Limitations of Current Approaches}

Realizing the theoretical gains of dynamic scheduling requires accurate, real-time detection of performance drifts. Static collective schedule synthesizers rely on active probing \ie{} injecting dummy traffic, which is effective offline but slows live training by consuming valuable network resources. Furthermore, relying solely on bandwidth measurements taken from network links or switches is insufficient. While these metrics capture the state of the network fabric, they fail to account for the end-to-end complexity of the communication path. Indeed, traditional network-level measurements \cite{molero22fancy,wu2025greyhound,dong2025sieve} overlook critical intra-node factors, such as PCIe bottlenecks and protocol overheads. Consequently, \gls{mccl} must explicitly monitor the full GPU-to-GPU data paths. While recent frameworks like MCCS~\cite{Yongji2024mccs} attempt runtime adaptability, their scope is limited to re-ordering ranks within a fixed ring algorithm; they lack the mechanisms to pinpoint path-specific degradations or the capability to synthesize new schedules online.
Moreover, accurate detection alone is not sufficient to enable dynamic schedule adaptation, as existing synthesizers fundamentally lack runtime architectural support. While synthesis tools like ForestColl~\cite{zhao2025forestcoll} are faster than traditional solver-based ones, they still operate entirely statelessly. When a localized gray failure occurs, they do not preserve unaffected portions of a previously optimal schedule. Instead, they discard the entire communication structure and rebuild from scratch, forcing redundant exploration and unnecessarily delaying runtime adaptation. For complex patterns like \textsc{AllToAll}, where every node sends unique data to every other node, the optimization challenge becomes intractable for time-discretized flow-based methods. While recent approaches like SyCCL \cite{jiamin2025syccl} attempt to mitigate this scaling issue by exploiting topology symmetry, they are rendered ineffective by gray failures, which inherently break symmetry and create irregular network states. Concurrent work OptCCL~\cite{shapley2026optccl} faces the same limitation: although its \gls{lp} is polynomial, its scalability relies on structural symmetry~\cite[Appendix~A.11]{optccl-tr}. Consequently, finding an optimal, asymmetry-aware \textsc{AllToAll} schedule for a large cluster remains computationally prohibitive for online adaptation. 


\section{System Design}

\label{sec:system}
\begin{figure}[t!]
    \begin{minipage}[c]{0.33\linewidth}
    \centering
        \includegraphics[width=0.89\linewidth]{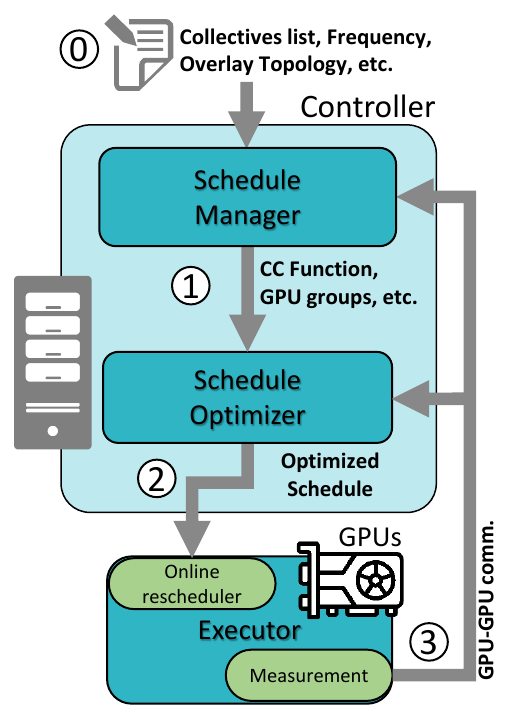}
        \caption{\gls{mccl} System architecture.}
        \label{fig:mccl}
    \end{minipage}
    \hfill
    \begin{minipage}[c]{0.64\linewidth}
    \centering
        \tiny
        \hrule height 0.6pt
        \vspace{-3.5pt}
        \captionof{algorithm}{Re-optimization selection algorithm.}
        \label{alg:prioritizecollectives}
        \vspace{-1.2pt}
        \hrule height 0.4pt
        \vspace{0.8pt}
            
            \begin{algorithmic}[1]
                \algtext*{EndIf}
                \algtext*{EndFor}
                \algtext*{EndWhile}

                \Require Graph $G(V,A)$, Measurements $\mathcal{M}$, Collectives $\mathcal{S}$ with active schedules, Bottleneck cuts $\{\mathcal{K}_s\}_{s \in \mathcal{S}}$, Threshold $\Gamma$, Base slacks $\{base\_slack_{s,a}\}_{s \in \mathcal{S},\, a \in \mathcal{K}_s}$

                \Statex \hspace{-0.4cm} \textbf{Definitions:} $\mathcal{C}(s)$: Commodities of collective $s$
                \Statex \hspace{0.8cm} $\mathcal{C}_a(s)$: Chunks of the active schedule for collective $s$ crossing link $a$
                \Statex \hspace{0.8cm} $\mathcal{K}_s$: Links saturated by the active schedule of $s$ (its bottleneck cut)
                \Statex \hspace{0.8cm} $x_{min}(s)$: Bottleneck data rate of the active schedule for collective $s$
                
                \For{each $s \in \mathcal{S}$} \Comment{--- For each collective and its active schedule ---}
                    \smallskip
                    \State $x_{min}(s) \gets \infty$ \Comment{\textbf{Phase 1:} Find Bottleneck Rate of $s$}
                    \For{each link $a \in A$} \Comment{Iterate all links in Topology}
                        \State $n_a \gets |\mathcal{C}_a(s)|$
                        \If{$n_a > 0$}
                            \State $x_{curr} \gets \mathcal{M}_a / n_a$
                            \State $x_{min}(s) \gets \min(x_{min}(s), x_{curr})$ \Comment{Update bottleneck of $s$}
                        \EndIf
                    \EndFor
                    \smallskip
                    \State $Slack_{max} \gets 0$ \Comment{\textbf{Phase 2:} Evaluate Slack on the Cut}
                    \For{each link $a \in \mathcal{K}_s$} \Comment{Iterate only links in the Cut}
                        \State $n_a \gets |\mathcal{C}_a(s)|$
                        
                        \State $u_{a} \gets n_a \times x_{min}(s)$
                        \State $current\_slack \gets (\mathcal{M}_a - u_{a}) / \mathcal{M}_a$ \Comment{Unused bandwidth on $a$}
                        \State $\delta_a \gets |current\_slack - base\_slack_{s,a}|$ \Comment{Deviation from rounding slack}
                        \State $Slack_{max} \gets \max(Slack_{max}, \delta_a)$
                    \EndFor
                    \smallskip
                    \If{ $Slack_{max} > \Gamma$ } \Comment{\textbf{Phase 3:} Decision}
                        \State $\text{ScheduleOptimizer}(\mathcal{C}(s), \mathcal{M})$ \Comment{New active schedule for $s$}
                    \EndIf
                \EndFor
                
            \end{algorithmic}
        
        \hrule height 0.4pt
    \end{minipage}
\end{figure}

The key idea of \gls{mccl} is to shift from a static ``optimize-once'' model to a dynamic ``always-optimizing'' framework. To achieve this without delaying the active training job, \gls{mccl} utilizes an asynchronous controller-executor architecture, depicted in \Cref{fig:mccl}. The \textbf{Controller} runs as a background service (\eg on a dedicated CPU or separate node) and manages the \textit{slow-path}: processing telemetry, diagnosing network health, and executing the schedule synthesis algorithms. The \textbf{Executor} runs on the GPUs participating in distributed training and manages the \textit{fast-path}: executing the collectives, passively measuring network throughput, and hot-swapping schedules.

By design \gls{mccl} targets communication-intensive operations (\textsc{AllGather}, \textsc{AllToAll}). However, distributed training paradigms such as \gls{fsdp} rely on operations like \textsc{AllReduce}. \gls{mccl} natively supports these workloads through collective decomposition. When a training job triggers an \textsc{AllReduce}, \gls{mccl} splits it into its constituent phases. The initial arithmetic-intensive \textsc{ReduceScatter} phase is executed using static schedules to leverage specialized in-network aggregation hardware~\cite{sharp}. The subsequent data-intensive \textsc{AllGather} phase is routed through \gls{mccl}'s Executor for dynamic optimization. This split preserves hardware-accelerated reduction performance while adapting the data-transfer phase to network conditions.

\subsection{The Dynamic Optimization Loop}
\label{sec:optimization_loop}
\gls{mccl} operates as a continuous feedback loop (\Cref{fig:mccl}), with a \textit{startup phase} for baseline optimization and a \textit{runtime phase} for dynamic adaptation. At startup \textcircled{0}, a profile of the training's collective calls (\eg list and frequency) and an initial overlay topology are fed to the \textit{Schedule Manager}. Such topology models communication as overlay links: each GPU connects to an intra-node switch (\eg{} NVSwitch), and GPUs in different nodes are connected by direct inter-node links. Initial link capacities are obtained via probing, as for other static synthesizers~\cite{cowan2023msccl,shah2023taccl,xuting2024teccl,zhao2025forestcoll}. In this phase, without online measurements, the \textit{Schedule Manager} \textcircled{1} acts as a static orchestrator, passing the topology to the \textit{Schedule Optimizer} to generate a set of optimized baseline schedules. These are sent to the Executor’s \textit{Online Rescheduler} \textcircled{2} to ensure that training starts with high-performance primitives. 
Optionally, \gls{mccl} can bypass this step and fall back to standard \gls{nccl} rings, deferring all optimization to the runtime phase.
Once training begins, the system enters the \textit{runtime phase} and transitions to its dynamic loop. While the optimized schedules are executed, the Executor's \textit{Measurement} module passively monitors overlay links throughput \textcircled{3}. This real-time data is continuously fed back to the Controller, fundamentally changing the \textit{Schedule Manager's} role. 

At runtime, the \textit{Schedule Manager} uses this feedback to diagnose the communication paths' health. It identifies performance drift (\eg due to gray failures) and dynamically selects which collectives require re-optimization. The \textit{Schedule Manager} then invokes the \textit{Schedule Optimizer} using the measurements from the GPUs as the new input topology. The resulting re-optimized schedules are sent to the \textit{Online Rescheduler} \textcircled{2}, which hot-swaps the algorithms. This continuous loop ensures that collective schedules evolve in lockstep with the underlying network's conditions. 


\subsection{Controller Components}
\label{subsec:controller_components}
The Controller operates on the ``slow-path'' and is composed of two modules:

\smallskip 
\noindent\textbf{Schedule Manager.} This component acts as the ``brain'' of the \gls{mccl} framework; its primary role is to select which collective(s) to re-optimize based on network conditions. It takes three inputs: the application profile with its graph $G(V,A)$ (step \textcircled{0}), the set of collectives $\mathcal{S}$ with the details of their active execution schedules, and throughput measurements $\mathcal{M}$ (step \textcircled{3}). Executors write raw throughput samples via RDMA into a dedicated per-node memory region on the Controller (\Cref{fig:matrixmeasure}, \Cref{app:measurements} for reference). The Controller applies an \gls{ewma} with smoothing factor $\alpha$ to these samples; the resulting smoothed throughput for each link $a$ is then stored as $\mathcal{M}_a$. This filtering explicitly attenuates short-lived microbursts, ensuring that only persistent degradations shift $\mathcal{M}$ enough to trigger a re-synthesis cycle. 

\Cref{alg:prioritizecollectives} details how the Schedule Manager selects the collectives to re-optimize. In a schedule, tensors are split into chunks, each routed along a path (\textsc{AllToAll}) or a tree (\textsc{AllGather}); $n_a$ denotes the number of chunks crossing link~$a$. At synthesis time, the Schedule Optimizer also returns the schedule's \textit{bottleneck cut} $\mathcal{K}_s$, i.e., the links the schedule saturates, and, for each link $a \in \mathcal{K}_s$, its $base\_slack_{s,a}$, i.e., the fraction of bandwidth left unused on $a$ because the solution is rounded into chunks. At runtime, the Schedule Manager checks whether, for each schedule $s$, $\mathcal{K}_s$ is still saturated under the current measurements $\mathcal{M}$. This takes three phases. \textit{Phase~1} computes the bottleneck rate $x_{min}(s) = \min_{a:\, n_a > 0} \mathcal{M}_a / n_a$, which represents the maximum per-chunk throughput allowed by the most congested link. \textit{Phase~2} computes, for each $a \in \mathcal{K}_s$, the current slack $(\mathcal{M}_a - n_a\, x_{min}(s)) / \mathcal{M}_a$ and its deviation $\delta_a$ from $base\_slack_{s,a}$. \textit{Phase~3} flags $s$ if $\max_{a \in \mathcal{K}_s} \delta_a$ exceeds a threshold $\Gamma$, and invokes the Schedule Optimizer with the commodities $\mathcal{C}(s)$ and the current $\mathcal{M}$. Both $\Gamma$ and the \gls{ewma}~$\alpha$ are tuned to the cluster profile: a small $\alpha$ filters out longer bursts at the cost of slower detection, while a large $\Gamma$ ignores milder degradations. Both reduce the number of re-optimizations and thus the risk of \textit{schedule thrashing}, \ie{} swapping schedules when degradations clear faster than a synthesis cycle. If a degradation clears during re-synthesis, the Schedule Manager aborts it and keeps the current schedule; if it clears right after, the installed schedule may be suboptimal until the next re-optimization. In practice, thrashing is unlikely, as gray failures typically persist for minutes to hours (\Cref{subsec:challenges}), far longer than a re-synthesis cycle. Additional thrashing prevention mechanisms and $\Gamma$ and $\alpha$ tuning from the observed failure dynamics are left for future work.
\smallskip
\noindent\textbf{Schedule Optimizer.} The optimizer serves as a unified wrapper that delegates collective optimization to specialized synthesizers. To ensure broad applicability while maintaining tractability, it targets communication-intensive primitives from which other collectives are derived: \textsc{AllToAll} and \textsc{AllGather}. For \textsc{AllToAll}, where asymmetric gray failures render traditional symmetry-based synthesizers ineffective, the optimizer invokes a \gls{cg}-based synthesizer. This engine enables fast re-synthesis via warm starts from previous solutions, supports time-bounded execution, and optimizes throughput against network irregularities. For \textsc{AllGather}, it interfaces with an extended ForestColl tree-packing synthesizer~\cite{zhao2025forestcoll}, which integrates an $A^*$-guided search to warm-start re-synthesis, preserve previous schedule structures, and minimize recomputation overhead. The underlying synthesis algorithms are detailed in \Cref{sec:optimization}.

\subsection{Executor Components} 

The Executor handles ``fast-path'' operations on the GPUs. Built on top of the MSCCL \cite{msccl,cowan2023msccl} backend to inject custom schedules, it consists of two elements:

\smallskip
\noindent\textbf{Online Rescheduler.} This module is responsible for the dynamic deployment of new schedules leveraging a runtime ``hot-swap'' mechanism adapted from MCCS \cite{Yongji2024mccs}. When the Controller provides a new Optimized Schedule (step \textcircled{2}), the Online Rescheduler stages it. At that point, GPUs belonging to the target communicator execute an out-of-band \textsc{AllGather} to exchange two state variables: the version number of their most recently staged schedule, and the sequence number of their last-launched collective (or collective group). Here, the minimum version number identifies the newest schedule staged across all GPUs, while the maximum sequence number defines the collective selected for the transition. Collectives with previous sequence numbers complete on the old schedule; subsequent ones complete using the re-optimized schedule. Because each swap is scoped to a single communicator, collectives on other communicators are unaffected.

\smallskip
\noindent\textbf{Measurement.} This module provides 
real-time telemetry feeding the optimization loop (step \textcircled{3}). To ensure that measurements capture only actual network communications, telemetry collection is restricted to operations without reduction phases, isolating link throughput from the computational phases of reduction arithmetic. To capture the distinct performance characteristics of different interconnects, the measurement module employs a hybrid metric strategy. For intra-node communication, it measures the cumulative per-GPU throughput, aggregating bandwidth across all concurrent transfers thus capturing the total device injection capacity. For inter-node communication, it captures point-to-point throughput between GPU pairs. These metrics are collected passively during training and asynchronously flushed to the Controller with RDMA\footnote{Telemetry offload leverages compute-bound phases or auxiliary NICs not used for model synchronization during training.}, which records them in a raw measurement matrix later used to derive $\mathcal{M}$ (\Cref{app:measurements}, \Cref{fig:matrixmeasure} for reference). The design of this passive measurement system is detailed in \Cref{sec:measurements}.

\section{Collectives Optimization}
\label{sec:optimization}
\subsection{Overview}
Optimizing collective communication centers on scheduling data-block (\ie{} tensor) transfers to minimize completion time. For \gls{mccl}, this requires balancing theoretical throughput optimality with the computational scalability required for online deployment. Existing synthesis tools fail to reconcile these conflicting goals because they cannot efficiently manage the optimization state during dynamic network performance drifts. Approaches like TECCL \cite{xuting2024teccl} optimize throughput-latency trade-offs but rely on computationally heavy formulations, such as time-expanded graphs or rigid \gls{milp}. These methods scale poorly with cluster size and may require substantial solving times, rendering them impractical for online schedule re-computation. Conversely, synthesizers like ForestColl~\cite{zhao2025forestcoll} are significantly faster at generating throughput-optimal schedules, but they 
operate statelessly. When a localized topology change occurs, these algorithms discard the entire communication structure and rebuild the schedule from scratch (\ie{} cold start) wasting critical computational cycles and delaying runtime adaptation. 

The core principle underlying \gls{mccl}'s optimization framework is that network anomalies typically disrupt only a localized fraction of the topology. Rather than re-evaluating the entire search space, \gls{mccl} utilizes the previously optimal solution as a structural heuristic to constrain the new search. The shift from a naive cold-start to an efficient warm-start paradigm drastically reduces redundant exploration. As our evaluations demonstrate, using historical schedules to anchor the search reduces convergence times, rendering real-time schedule adaptation practical. To achieve this, the Schedule Optimizer integrates two specialized synthesis engines tailored to the communication structures of \textsc{AllGather} and \textsc{AllToAll}:

\smallskip
\noindent
\textbf{\textsc{AllToAll} via \acrfull{cg}.}
We propose an \gls{lp}-throughput-optimal, decomposition-based formulation for \textsc{AllToAll} collectives that eliminates solving time overhead associated with traditional time-expanded graphs.
By decoupling the problem into a \gls{rmp} and a pricing subproblem, we avoid enumerating the exponentially many tree variables while preserving \gls{lp}-optimality. Building on this scalable foundation, we also introduce a warm-start mechanism that retains valid routing structures from previous iterations, drastically reducing the time required to adapt to network changes.

\smallskip
\noindent
\textbf{\textsc{AllGather} via ForestColl-warm.} We introduce ForestColl-warm, which extends the throughput-optimal tree-packing algorithm by incorporating an $A^*$-guided exploration strategy tailored for online re-computation. By leveraging the solution history to direct the underlying \gls{bfs}, this mechanism strategically reorders the sequence of candidate links to align with the previous optimal arborescence. This guided approach aggressively prunes the exploration space, minimizing redundant computational overhead and enabling rapid adaptation to network changes. Additionally, in \Cref{app:latency_switch_removal} we propose a ForestColl-like latency-sensitive formulation which reduces maximum tree depths with the objective of reducing schedules' latency.

\subsection{AllToAll optimization}
\label{subsec:a2a_optimization}

To satisfy the strict time requirements of online re-synthesis, we model the \textsc{AllToAll} scheduling problem as a throughput maximization problem, which we solve using a Dantzig--Wolfe reformulation \cite{dantzig-wolfe} with \acrfull{cg}. Let $\mathcal{T}_{c}$ be the set of all shortest-path trees rooted at the source of tensor $c \in C$. We define $z$ as the throughput, and $x_{T_{c}}$ as the volume of traffic sent along the shortest-path tree $T_{c} \in \mathcal{T}_{c}$ for a given $c \in C$. The optimization problem is formulated as follows:
%
%
\begin{smallalign}
&\max \quad z \span\span\span \label{a2a:obj}\\
\alpha_c:\ & \sum_{T_{c}\in \mathcal{T}_{c}} x_{T_{c}} \;\ge\; z
    & & \forall c\in C, \label{a2a:throughput}\\
\alpha_{a}:\ & \sum_{c\in C} \sum_{T_c\in \mathcal{T}_c:\,a\in T_c} \beta_a^{T_{c}}\,x_{T_c}
    \;\le\; c_{a}
    & & \forall a\in A, \label{a2a:capacity}\\
& x_{T_c} \ge 0
    & & \forall c\in C,\ \forall T_c \in \mathcal{T}_c. \label{a2a:bound}
\end{smallalign}
where $\beta_a^{T_{c}}$ is the number of shortest paths within the tree $T_c$ that use link $a$. The objective~\eqref{a2a:obj} maximizes the collective throughput $z$. Constraints~\eqref{a2a:throughput} are throughput constraints: they tie each tensor's total routed flow to $z$, so that maximizing $z$ drives up the minimum per-tensor throughput (a max--min objective). Constraints~\eqref{a2a:capacity} are link-capacity constraints, ensuring the aggregate load on each link $a$ stays within its bandwidth $c_a$. Finally, constraints~\eqref{a2a:bound} enforce non-negativity of the flow variables. We denote the dual variables associated with~\eqref{a2a:throughput} and~\eqref{a2a:capacity} as $\alpha_c$ and $\alpha_a$, respectively.

This formulation is the Dantzig–Wolfe reformulation of a compact, arc-based multi-commodity flow model; we defer the compact model and its full derivation to \Cref{app:a2a_derivation}. Directly solving the compact model is computationally prohibitive at scale. The \gls{cg} approach overcomes this challenge by generating columns on the fly. In particular, the algorithm alternates between solving the \gls{rmp}, limiting the optimization to a subset of variables, and solving a pricing subproblem for each tensor. The pricing subproblem is parameterized by the dual variables $\alpha_c$ and $\alpha_a$ and identifies whether any variable not yet generated can improve the objective value; specifically, it searches for a shortest-path tree $T_{c}$ for each $c\in C$ such that $\sum_{a\in T_c} \beta_a^{T_{c}} \alpha_a < \alpha_c$. Intuitively, $\alpha_a$ duals act as congestion prices, so the left-hand side measures the congestion that $T_c$ would experience, while $\alpha_c$ is the marginal value of increasing the throughput of tensor $c$.

To account for the sequential nature of tensor transfers within the \gls{cg} framework, the pricing subproblem's shortest-path tree search is executed over a time-expanded graph $\tilde{G}=(\tilde{V}, \tilde{A})$, where \emph{steps} model the order of discrete chunk transfers between GPUs. To govern this expansion, we bound the maximum number of steps by a parameter $m_s$. The set of vertices $\tilde{V}$ consists of the original nodes $V$ replicated $m_s$ times, where the $i$-th copy of a node $v \in V$ is denoted $v_i \in \tilde{V}$. The set of arcs $\tilde{A}$ comprises two types of links to model intra-switch and inter-device propagation: for each link $(u,v)$ towards a switch node $v$, we add instantaneous links $(u_i, v_i)$ for $i \in \{0, \dots, m_s-1\}$; for all other links $(u,v)$ where $v$ is not a switch node, we add forward-stepping links $(u_i, v_{i+1})$ for $i \in \{0, \dots, m_s-2\}$. Since $\tilde{G}$ is acyclic by construction, the pricing reduces to a shortest-path tree search that we solve with a DAG-tailored 
algorithm processing $\tilde{V}$ in topological order (\ie{} step), which naturally favors shallow trees and runs in $O(|\tilde{V}|+|\tilde{A}|)$ time. This extended topology structurally limits any discovered path to at most $m_s$ steps and natively encourages the shortest-path algorithm to select the routes requiring the fewest steps among equal-cost alternatives. To prioritize the aggressive computational speeds required for online re-computation and warm-starting, we statically fix $m_s = |V| - 1$. While this constant bound guarantees maximum throughput with minimal solver overhead, the formulation can optionally minimize $m_s$ via a binary search if tighter latency bounds are desired. We defer the description of this extended formulation to \Cref{app:a2a_minsteps}.

Once the CG converges to the fractional LP-optimal solution, we apply deep-dive rounding~\cite{HUIN202372} to extract a schedule with at most ten chunks. This heuristic incurs a marginal gap to the \gls{milp} optimum, yielding a near-optimal executable schedule (evaluated in Section~\ref{subsec:algorithm-evaluation}, Appendix~\ref{app:a2a_rounding}).

\subsection{AllGather optimization}

In this section, we present our ForestColl extension, whose baseline algorithm comprises three phases that must be executed sequentially:

\smallskip
\noindent
\textit{Phase 1: Compute maximum throughput} --- This phase calculates the maximum flow for each tensor from a super-source $S$ to the source node of each tensor $c$ within the reversed graph. The overall maximum throughput is determined by the minimum of these computed flows.

\smallskip
\noindent
\textit{Phase 2: Switch removal} --- This phase abstracts switches by replacing physical paths with direct logical links. 
The procedure greedily assigns capacity to each logical link (Fig.~\ref{fig:FCPhyLog}, Appendix~\ref{app:latency_switch_removal}). For each node, the total capacity assigned to incoming and outgoing logical links must not exceed the underlying physical link capacity and must satisfy the calculated maximum throughput.


\smallskip
\noindent
\textit{Phase 3: Tree generation} --- For every tensor in the logical topology, this procedure allocates link capacity based on the maximum throughput. The allocation uses a \acrfull{bfs} strategy initiated at the tensor's source node, which guarantees an arborescence structure.

\smallskip
\noindent
\textbf{History-Guided Tree Generation (ForestColl-warm).} To guarantee the rapid schedule re-synthesis required for online adaptation, we enhance the tree generation phase by making it history-aware. The original ForestColl algorithm employs a stateless \gls{bfs} to construct the arborescence. This approach is computationally wasteful during a localized network anomaly, as it discards the entire communication structure and rebuilds from a cold start.

To improve the re-optimization strategy, our enhanced \gls{bfs} is guided by the previously computed arborescence through a mechanism analogous to the $A^*$ algorithm. The primary objective is to preserve its structure as much as possible. This approach aggressively reduces redundant explorations within the \gls{bfs} by reordering the sequence of candidate links. Given that a gray failure typically impacts a limited number of links, there is a high probability that links from prior optimal trees remain valid in subsequent iterations, allowing the solver to rapidly converge on a new throughput-optimal schedule.  

\smallskip
\noindent
\textbf{Balancing Latency and Computation Time.}
While the proposed history-guided search is designed to reduce schedule re-synthesis time, ForestColl's default greedy switch removal (Phase 2) ignores tree depth, which can lead to sparse topologies and deep communication trees, hence high latency. If rescheduling-time constraints can be relaxed and lower latency is preferred (e.g., for workloads dominated by smaller transfers), a step-aware switch-removal formulation can be used instead, which reduces maximum tree depth as a proxy for latency. We detail this extended formulation in \Cref{app:latency_switch_removal}.

\section{Measurements framework}
\label{sec:measurements}


\begin{figure}
    \begin{minipage}[b]{0.49\linewidth}
        \includegraphics[width=0.95\linewidth]{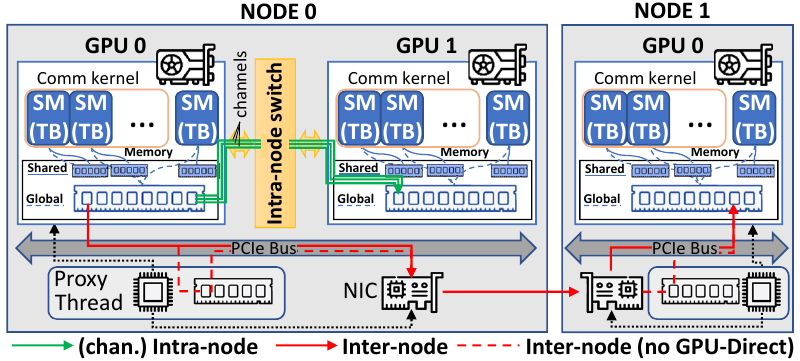}
        \caption{GPU Communication and Memory Layout.}
        \label{fig:gpu-comm}
    \end{minipage}
    \begin{minipage}[b]{0.49\linewidth}
        \includegraphics[width=\linewidth]{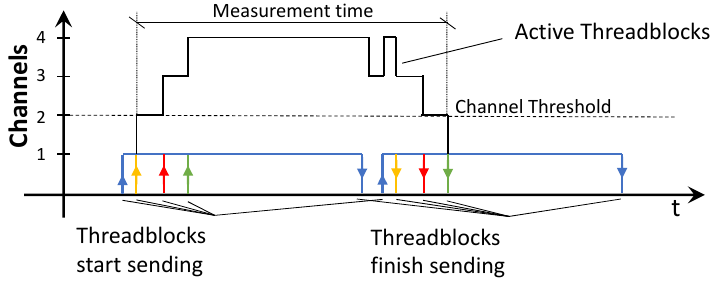}
        \caption{Per channel (TB) measurements process.}
        \label{fig:meas-proc}
    \end{minipage}
\end{figure}

Before detailing our lightweight passive measurement architecture, we analyze the execution model of the underlying communication library. Without loss of generality, we focus on NCCL, though other software stacks for GPU-to-GPU communication share fundamentally similar architectures. 

As detailed in recent literature \cite{hu2025demystifying}, NCCL parallelizes collective operations by splitting them into virtual pipelines known as \textit{channels}.
Each channel manages a fraction of the total payload via a producer-consumer model, transferring data from a source to a destination. At the hardware level, each channel is driven by a single \gls{tb} executing on a dedicated \gls{sm}. For throughput-intensive protocols (\eg{} Simple), NCCL partitions the payload across these channels. This enables fine-grained pipelining where computation, transmission, and synchronization occur concurrently. The underlying data movement relies on two distinct mechanisms: \textit{intra-} and \textit{inter-node} as depicted in  \Cref{fig:gpu-comm}. For intra-node communication, NCCL utilizes the \texttt{P2P\_DIRECT} transport. By mapping peer memory into a unified address space, \gls{tb} issue direct Load/Store instructions to remote GPU memory, bypassing intermediate staging buffers. Inter-node communication instead relies on a Proxy-based protocol run by the Host CPU. The GPU posts work requests to a shared FIFO in host memory, which the Proxy polls to trigger network operations (\eg{} RDMA or Socket transfers). With GPU-Direct, chunks are transferred immediately from the GPU to the \gls{nic}, bypassing the CPU for the payload. 

This decoupled, multichannel execution model presents unique challenges for telemetry: because traffic is split across independent hardware and channels, traditional single-point monitoring is insufficient. This motivates our specialized measurement architecture, which is designed to capture effective bandwidth natively across these distributed execution paths. Currently, our measurement framework targets collective operations without a reduction phase (\eg{} \textsc{AllGather}, \textsc{AllToAll}) to ensure that the measured throughput is not affected by the computational latency.


\subsection{Intra-node measurements}
\label{sec:intra-node-measurements}

\begin{figure}
    \hfill
    \begin{minipage}[b]{0.40\linewidth}
        \centering
        \includegraphics[width=0.9\linewidth]{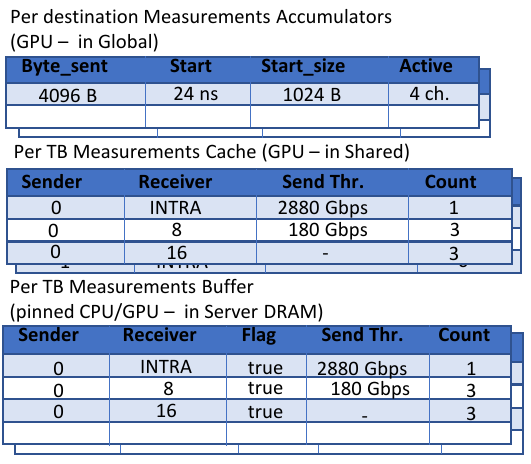}
        \caption{Measurements data structures.}
        \label{fig:datastructs}
    \end{minipage}
    \hfill
    \begin{minipage}[b]{0.59\linewidth}
        \centering
        \includegraphics[width=0.99\linewidth]{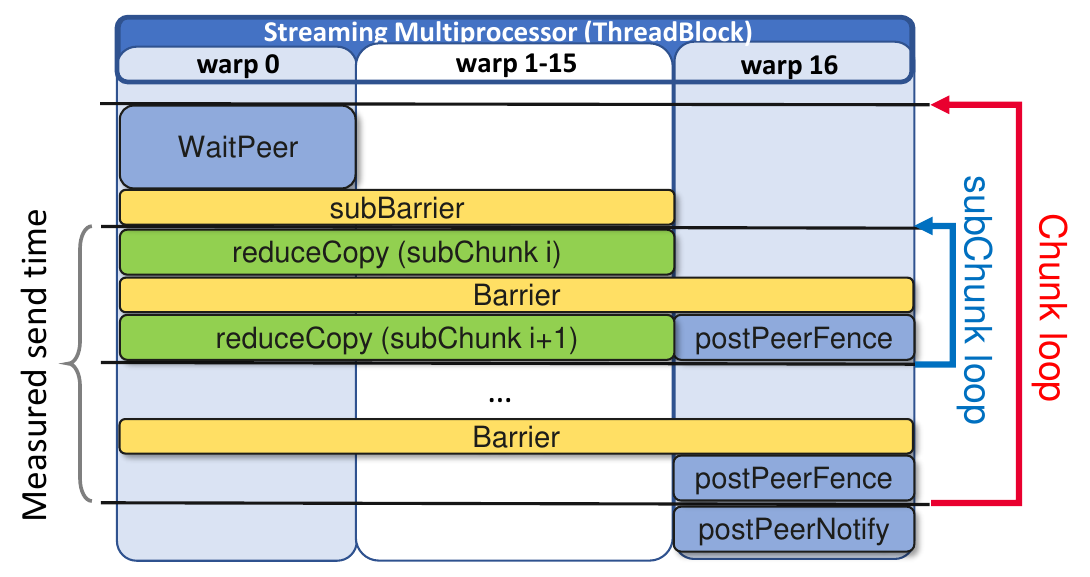}
        \caption{NCCL communication kernel schema.}
        \label{fig:warp-sched}
    \end{minipage}
    \hfill
\end{figure}

Measuring intra-node throughput presents two main challenges: memory contention and transient effects (naive measurements might underestimate throughput, including transient phases). To mitigate this, we introduce a conditional measurement algorithm supported by a three-tiered memory hierarchy. As illustrated in \Cref{fig:datastructs}, our measurement module uses three data structures:

\noindent
\textit{Measurements Cache} --- A per-\gls{tb} cache located in the fast Shared Memory. It aggregates bandwidth statistics \ie{} \texttt{count}, \texttt{sum\_bw}, to minimize DMA transfer from the GPU to the Host.

\noindent
\textit{Measurements Accumulator} --- A per-destination synchronization point stored in GPU global memory and accessible by all \gls{tb}s. It tracks the global \texttt{active} channel count to detect link saturation, and \texttt{bytes\_sent}, the cumulative amount of bytes transmitted to the destination. Additionally, it records \texttt{start} and \texttt{start\_size}, the time and byte count at the beginning of the measurement window. 

\noindent
\textit{Measurements Buffer} --- A pinned memory buffer allocated in DRAM to export data from the GPU to the Host. To avoid race conditions, each \gls{tb} is assigned a set of dedicated slots arranged as a circular buffer; if no slot is available, the measurement is dropped. This structure mirrors the fields of the cache but adds a \texttt{flag}, which signals to the Host that the 
data is valid for reading.

\smallskip
\noindent
\textbf{Measurement Algorithm.} To ensure correct measurements, we employ a protocol that records data only when the number of active channels exceeds a specific threshold. This threshold can be established in two ways. The primary method leverages NCCL’s native topology discovery, which derives optimal allocations to maximize parallelism. Alternatively, while NCCL’s allocations are robust, an empirical lower bound can be identified through offline profiling. Note that this static threshold does not need to be adapted in case of reduced bandwidth as the collective algorithm will always use the expected number of channels needed to saturate the nominal bandwidth.

The process, depicted in \Cref{fig:meas-proc}, follows this logic: as a \gls{tb} begins transmission on a channel, it atomically increments the \texttt{active} counter in the Measurements Accumulator. When \texttt{active} exceeds the defined channel threshold $N$ (cf. \Cref{fig:meas-proc}), the system enters the measurement state and the current \gls{tb} logs the \texttt{start} timestamp and current \texttt{bytes\_sent}. As channels complete their work, they decrement the \texttt{active} counter. When \texttt{active} count drops below the threshold, the measurement window closes. The \gls{tb} triggering this event calculates the bandwidth ($\mathit{BW} = \frac{\Delta \text{Size}}{\Delta \text{Time}}$) 
and stores it in the local Measurements Cache.\footnote{Note that in our prototype, a flag signals the state of the measurement accumulator \ie{} measuring, below\_threshold, and finalizing, to prevent a new measurement window from starting before the previous one is finalized.}


\smallskip
\noindent
\textbf{Implementation efficiency.}
To isolate chunk transfer times in the NCCL communication kernel (cf. \Cref{fig:warp-sched}), we exclude the initial \texttt{waitPeer} and \texttt{subBarrier} routines that await data readiness. Moreover, to ensure fine-grained data persistence without stalling primary data movers, we restructure the standard NCCL \gls{tb} signaling mechanism. In the original implementation, a single monolithic \texttt{postPeer} routine handles both memory consistency and peer notification. We split this operation into two distinct primitives: \texttt{postPeerFence}, which enforces memory ordering, and \texttt{postPeerNotify}, which signals completion. This separation allows us to reuse the dedicated signaling warp 16 as a progress bookkeeper, feeding the \gls{mccl} measurement data structures. By executing \texttt{postPeerFence} in parallel with the \texttt{reduceCopy}\footnote{In NCCL kernel, the \texttt{reduceCopy} symbol is immutable; for non-reduction operations it does not incur arithmetic overhead.} loop (warps 1–15), we enforce system-level memory fences on sub-block transfers without blocking the primary data movers. This effectively masks the latency of synchronization and atomic operations. Our evaluation shows that this instrumentation incurs a negligible overhead of $\approx$\SI{2}{\percent} provided the \gls{tb} allocation is sufficient to saturate the available bandwidth, a prerequisite for achieving throughput-optimal schedules.

\subsection{Inter-node measurements}

For inter-node communication, we leverage the CPU Proxy Thread to capture performance metrics. Unlike the highly parallelized execution model of GPU \gls{tb}s, the Proxy centralizes the management of network operations for a given connection. This significantly simplifies the measurement logic and effectively eliminates the need for atomic operations. Despite these architectural differences, we reuse the same \textit{Measurements Accumulator} and \textit{Measurements Buffer} data structures, as well as the same multi-channel measurement algorithm. The primary distinction is that, in practice, a single communication channel is often sufficient to saturate the available network bandwidth. 

The measurement logic is embedded within the Proxy's Finite State Machine (FSM), which manages the life-cycle of network chunks in three stages: \textit{Posted}--a transfer slot is available, \textit{Transmitted}--the network operation, \eg{} \texttt{isend}, has been triggered, and \textit{Done}--transmission is confirmed as complete. To calculate the effective bandwidth, we isolate network injection and transit time by measuring the interval between \textit{Transmitted} and \textit{Done} states. To ensure accuracy, \gls{mccl} models this effective inter-node throughput as the minimum of the local buffer write rate and the network travel rate (\ie{} local-to-remote buffer transfer), thereby accounting for potential bottlenecks across the entire data path, i.e., \gls{hbm}, PCIe bus, or network.

\section{Experimental Results}
\label{sec:results}



We validate \gls{mccl} at three different levels. First, we evaluate our optimization algorithms against state-of-the-art competitors in terms of 
solving time
and schedule performance when optimizing a single collective in offline settings. Second, we conduct a closed-loop evaluation on a prototype of \gls{mccl} to assess the overhead and accuracy of the measurement module, as well as the online rescheduling mechanism. Finally, we run large-scale simulations of real-life training workloads to quantify end-to-end advantages of \gls{mccl} online optimization in terms of training duration.


\subsection{Optimization Algorithm}
\label{subsec:algorithm-evaluation}

We evaluate solving time and schedule performance against TECCL and vanilla ForestColl, varying payload and group size. For solving-time experiments, we use a 32-core server with a commercial solver, TECCL code~\cite{teccl}, and our C++ re-implementation of the released ForestColl~\cite{forestcoll}. We evaluate all synthesizers with up to ten chunks and report each time the best-performing configuration.\footnote{Due to prohibitive solving times, we limit TECCL to a single chunk (two when specified) when solving \textsc{AllGather} collectives; for each run we report the fastest of its Optimal and Heuristic variants.}
To evaluate schedule performance, we run simulations on a modified version of SimAI~\cite{wang2025simai,simai} that supports custom CCL execution.\footnote{https://github.com/NetMeasurements-Team/SimAI.} We consider HPN rail optimized topologies \cite{kun2024hpn}, featuring 8 GPUs per host, each with a dedicated NIC. Intra-node communication relies on an NVSwitch (\SI{2880}{\Gbps}, \SI{25}{\nano\second} delay), while inter-node traffic traverses Top-of-Rack (ToR) switches (\SI{400}{\Gbps}, \SI{0.5}{\micro\second} delay). The topology is composed of multiple segments, each containing 8 ToR switches and up to 32 hosts (256 GPUs). Aggregation switches interconnect these segments in a single-plane configuration via \SI{400}{\Gbps} links. In the following, we denote by \texttt{S$\times$H$\times$N} a topology of \texttt{S} segments, \texttt{H} hosts per segment, and \texttt{N} GPUs per host. 

\noindent\textbf{Failures.}
We define three different failures to evaluate our approach under diverse conditions. \textit{Failure 1} (Fail1) affects the NVSwitch of the first host; we simulate that 13 of the 18 lanes connecting GPU 0 fail, and the intra-node bandwidth drops from \SI{2880}{\Gbps} to \SI{800}{\Gbps}. In \textit{Failure 2} (Fail2), we reduce the capacity of the link connecting GPU 0 to its ToR switch from \SI{400}{\Gbps} to \SI{50}{\Gbps}. \textit{Failure 3} (Fail3) models a scenario where GPU 0 fails and is replaced by a backup GPU in a different rack, incurring extra \SI{1}{\micro\second} of ToR switch latency. To accommodate the traffic distribution, where the bulk of the load remains with the original node, the backup is provisioned with a dedicated \SI{50}{\Gbps} bandwidth to the faulty node, while communications to the wider data center share \SI{100}{\Gbps}.

\begin{figure}
    \centering
    \begin{subfigure}[b]{0.245\linewidth}
        \includegraphics[width=\linewidth]{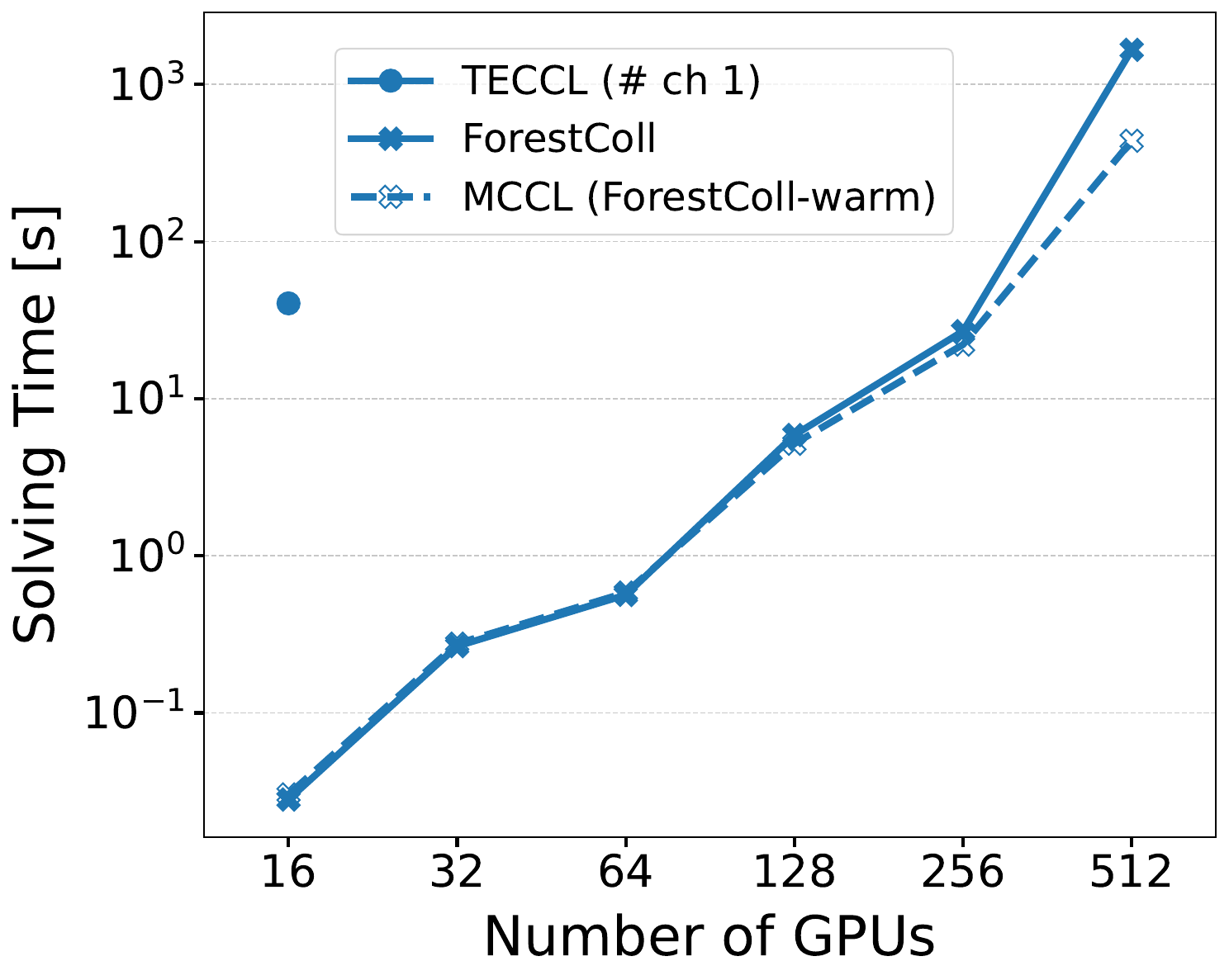}
        \caption{\textsc{AllGather}, Fail 1.}
        \label{fig:solving_timeAGF1}
    \end{subfigure}
    \hfill
        \begin{subfigure}[b]{0.245\linewidth}
        \includegraphics[width=\linewidth]{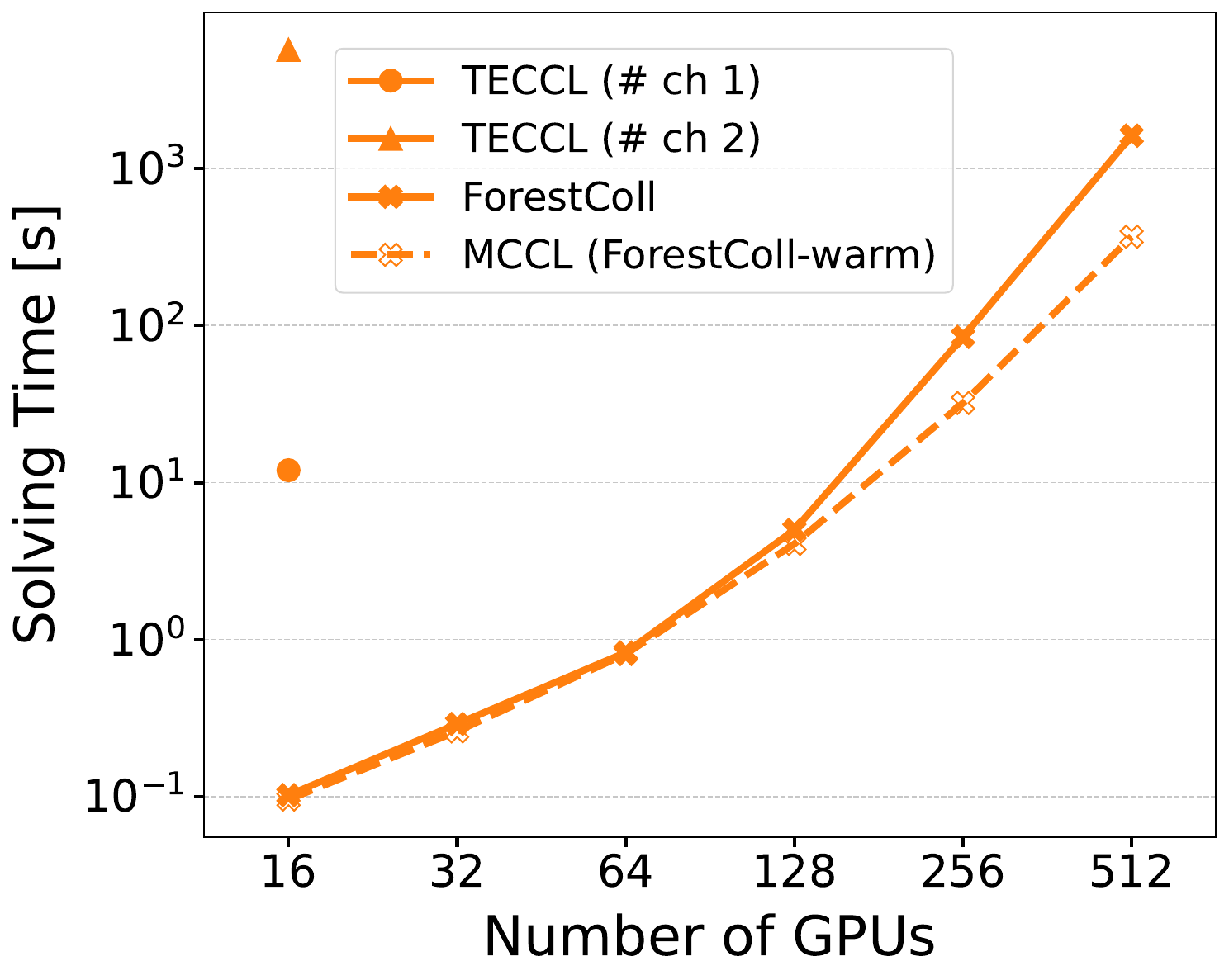}
        \caption{\textsc{AllGather}, Fail 2.}
        \label{fig:solving_timeAGF2}
    \end{subfigure}
    \hfill
    \begin{subfigure}[b]{0.245\linewidth}
        \includegraphics[width=\linewidth]{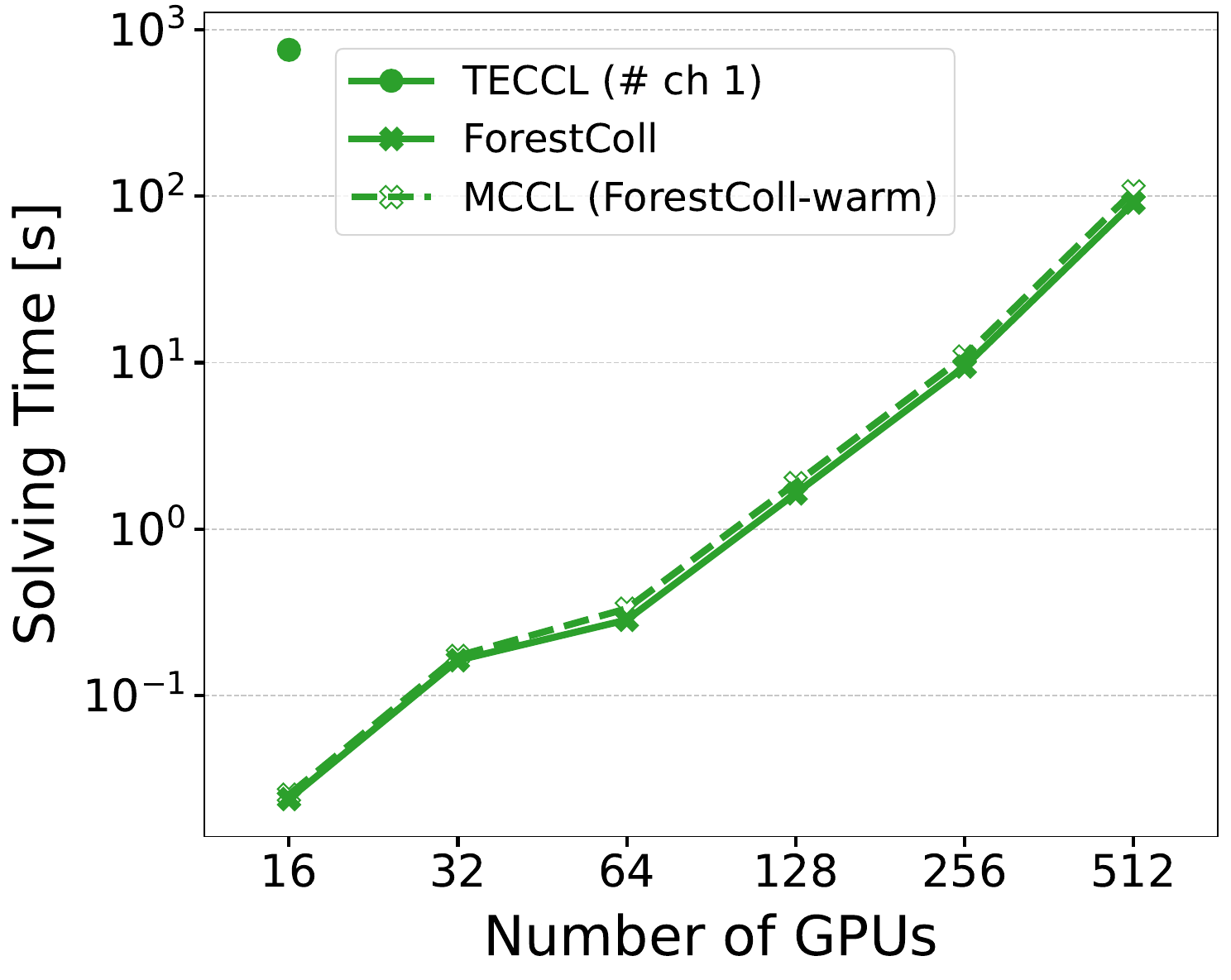}
        \caption{\textsc{AllGather}, Fail 3.}
        \label{fig:solving_timeAGF3}
    \end{subfigure}
    \hfill
    \begin{subfigure}[b]{0.245\linewidth}
        \includegraphics[width=\linewidth]{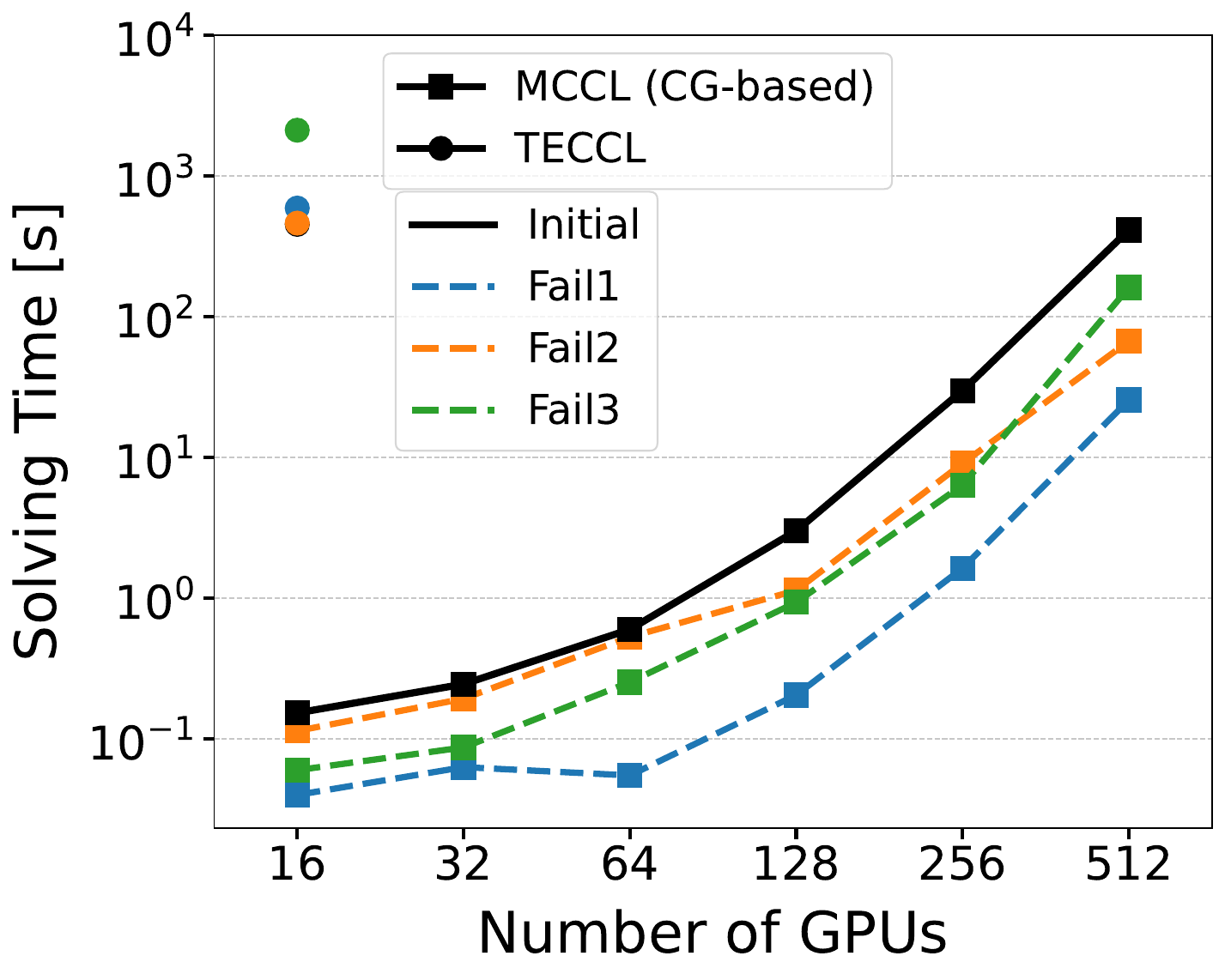}
        \caption{\textsc{AllToAll}, Fails 1-3.}
        \label{fig:solving_timeA2A}
    \end{subfigure}
    \caption{Optimization solving time comparison for initial schedules and re-optimized one in case of failure.}
    \label{fig:solving_time}
\end{figure}

\smallskip
\noindent
\textbf{Solving Time.}
We evaluate the scalability of MCCL's synthesizers against TECCL~\cite{xuting2024teccl} and vanilla ForestColl~\cite{zhao2025forestcoll} varying the cluster size.  
Figure~\ref{fig:solving_time} illustrates the solving times for \textsc{AllGather} (\Cref{fig:solving_timeAGF1,fig:solving_timeAGF2,fig:solving_timeAGF3}) and \textsc{AllToAll} (\Cref{fig:solving_timeA2A}) under the previously described failures.
Across all configurations, TECCL proves computationally intractable even for small topologies (up to \textbf{\SI[print-unity-mantissa=false]{1e3}{\second}} for a 16-GPU \textsc{AllToAll}) as also reported in \cite{zhao2025forestcoll,optccl-tr}. In contrast, ForestColl and our CG-based algorithm scale significantly better, computing schedules for collectives of up to 512-GPUs. Regarding \textsc{AllGather} collectives, our version of ForestColl enhanced with warm-start is up to $\mathbf{3.79{\times}}$ and $\mathbf{4.40{\times}}$ faster than the base algorithm for Failures 1 and 2, for which the underlying graph structure is highly reusable. Warm-start yields instead no noticeable improvement for Failure 3, which is significantly more disruptive than Failures 1 and 2 and requires recomputation of a great part of the graph. Finally, \Cref{fig:solving_timeA2A} shows that our CG-based algorithm greatly benefits from the warm-start mechanism. It re-optimizes \textsc{AllToAll} collectives in a fraction of the initial solving time (up to $\mathbf{18.3{\times}}$ faster) and provides solutions within tens of seconds for 512-GPUs collectives. For this collective family, warm-start is up to $\mathbf{17.8{\times}}$ faster than recomputing from scratch (result not shown).


\begin{figure}[!t]
    \centering
    \begin{subfigure}[b]{0.49\linewidth}
        \includegraphics[width=\linewidth]{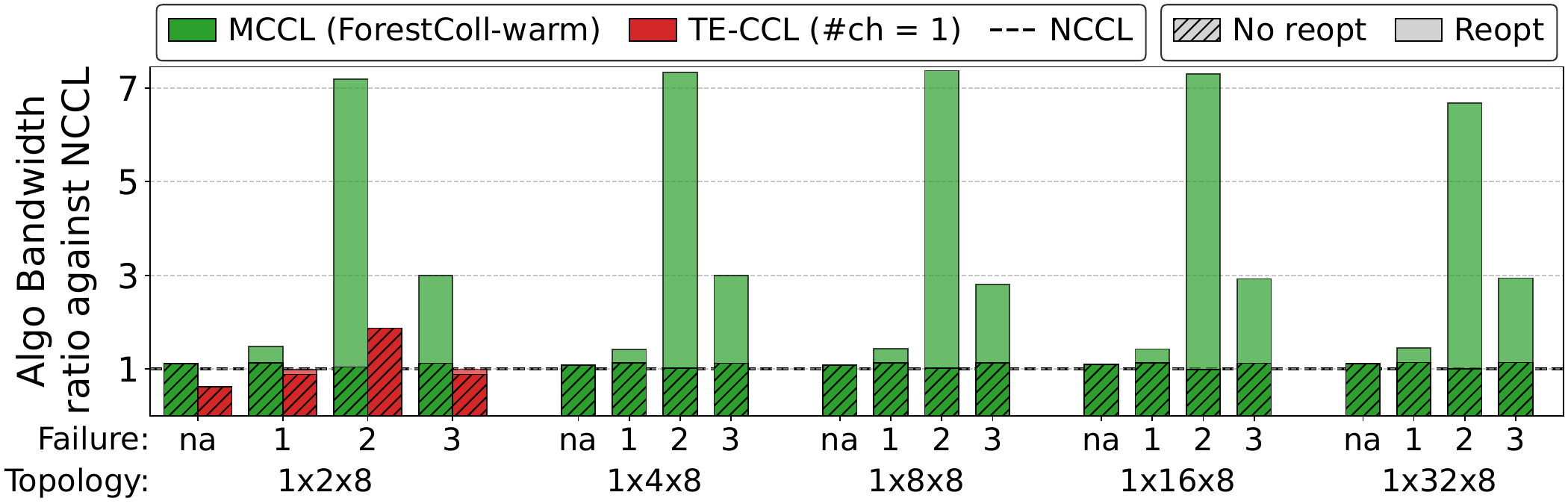}
        \caption{\textsc{AllGather} performance against NCCL.}
        \label{fig:ag_quality}
    \end{subfigure}
    \hfill
    \begin{subfigure}[b]{0.49\linewidth}
        \includegraphics[width=\linewidth]{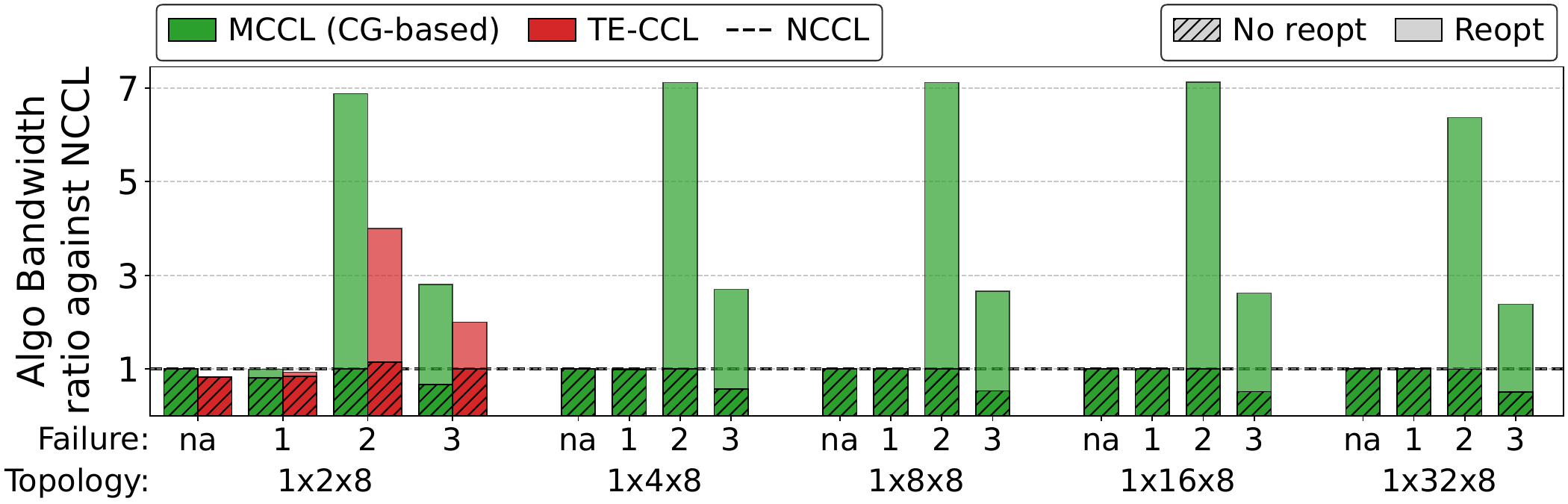}
        \caption{\textsc{AllToAll} performance against NCCL.}
        \label{fig:a2a_quality}
    \end{subfigure}
    \caption{Schedule performance for \textsc{AllGather} (a), and \textsc{AllToAll} (b).}
    \label{fig:quality}
\end{figure}

\newcommand{\maxError}{2.13}

\smallskip
\noindent
\textbf{Schedule Performance.}
To quantify the throughput gains offered by \gls{mccl}, we conducted offline simulations on our extended version of SimAI with the NS-3 network backend.
Using a \SI{128}{\mega\byte} tensor size per source, we measured the \textit{Algorithm Bandwidth}~\cite{algobw} speedup relative to NCCL.  As detailed in \Cref{subsec:a2a_optimization}, a rounding phase is required for the \textsc{AllToAll} collective, introducing a maximum error of \textbf{\SI{\maxError}{\percent}} in our experiments with respect to the optimal. TECCL time-expanded formulation enforces link capacities per epoch, making this conversion inherently harder than our column-generation rounding. This accounts for a gap with respect to \gls{mccl} even with no failures, despite using the same chunk budget. Results on \gls{mccl} rounding error are reported in \Cref{app:a2a_rounding}.

Figure~\ref{fig:quality} shows the performance of \textsc{AllToAll} and \textsc{AllGather} under the failure scenarios defined above: intra-node degradation, inter-node bottleneck, and GPU replacement. Results highlight the brittleness of static approaches, \ie{} ``No reopt''. When physical network conditions deviate from pre-computed topology assumptions, their performance collapses to baseline levels ($\mathbf{{\approx}1.0\times}$). In contrast, \gls{mccl} adapts effectively to varying link capacities. For \textsc{AllGather} (cf. Figure~\ref{fig:ag_quality}), \gls{mccl} achieves speedups of up to $\mathbf{7.33\times}$ over the baseline in the presence of gray failures. For \textsc{AllToAll} (cf. Figure~\ref{fig:a2a_quality}), \gls{mccl} significantly outperforms static synthesizers with speedups of up to $\mathbf{7.12\times}$. It is important to note that \gls{mccl} optimizes for computational efficiency. As detailed in \Cref{subsec:controller_components}, the system employs a cut-based filtering mechanism. For this reason, when failures occur on links outside the critical cut of the active collective, or when a cut is uniformly scaled down, the existing schedule remains throughput-optimal. In these cases, \gls{mccl} correctly identifies that the slack is minimized and suppresses re-optimization, thereby conserving precious computational resources.


\subsection{MCCL Prototype}

We evaluate a prototype implementation of \gls{mccl} that builds on MSCCL \cite{msccl}. The prototype runs on two hosts, each featuring four Nvidia H200 GPUs connected in a rail-optimized topology with Nvidia ConnectX-7 NICs (\SI{200}{\Gbps}) using GPU-Direct (details in \Cref{tab:prototype}).

\begin{figure}
\hfill
\begin{minipage}[b]{0.545\linewidth}
\captionsetup{type=figure}
    \vspace{-1.5pt}
    \centering
    \begin{subfigure}{0.49\linewidth}
        \includegraphics[width=\linewidth]{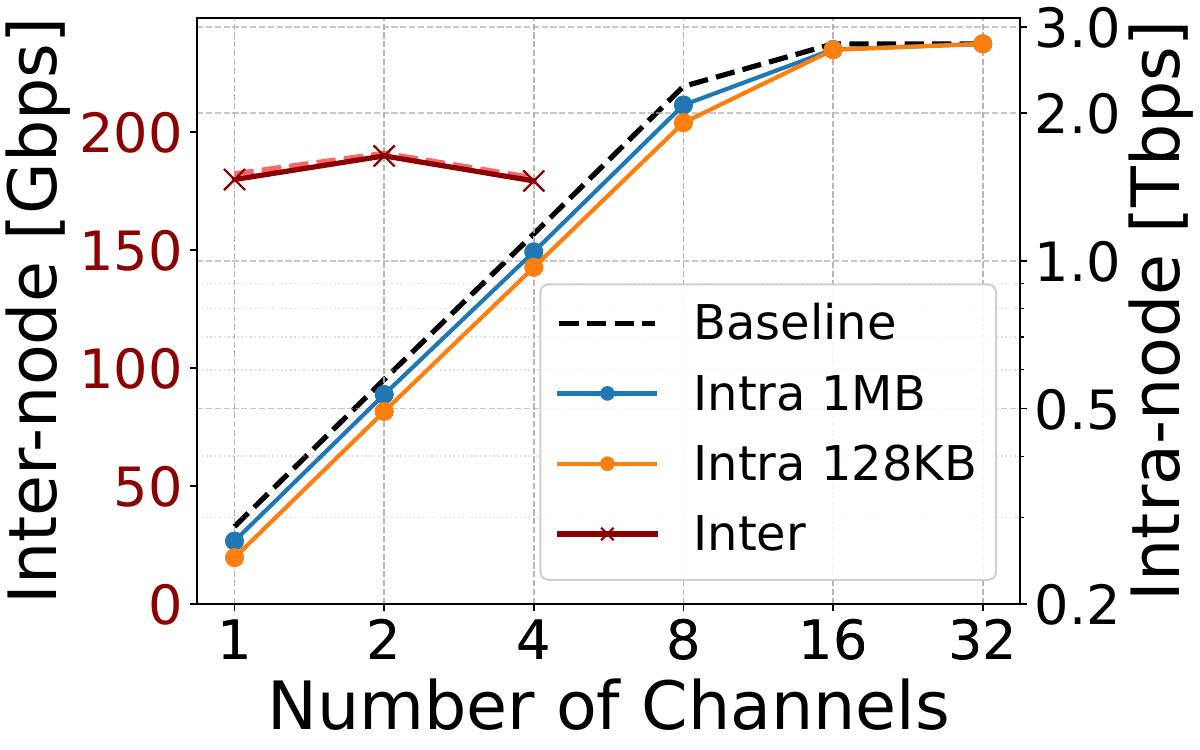}
        \caption{Throughput overhead.}
        \label{fig:bandwidth_overhead}
    \end{subfigure}
    \hfill
    \begin{subfigure}{0.49\linewidth}
        \includegraphics[width=\linewidth]{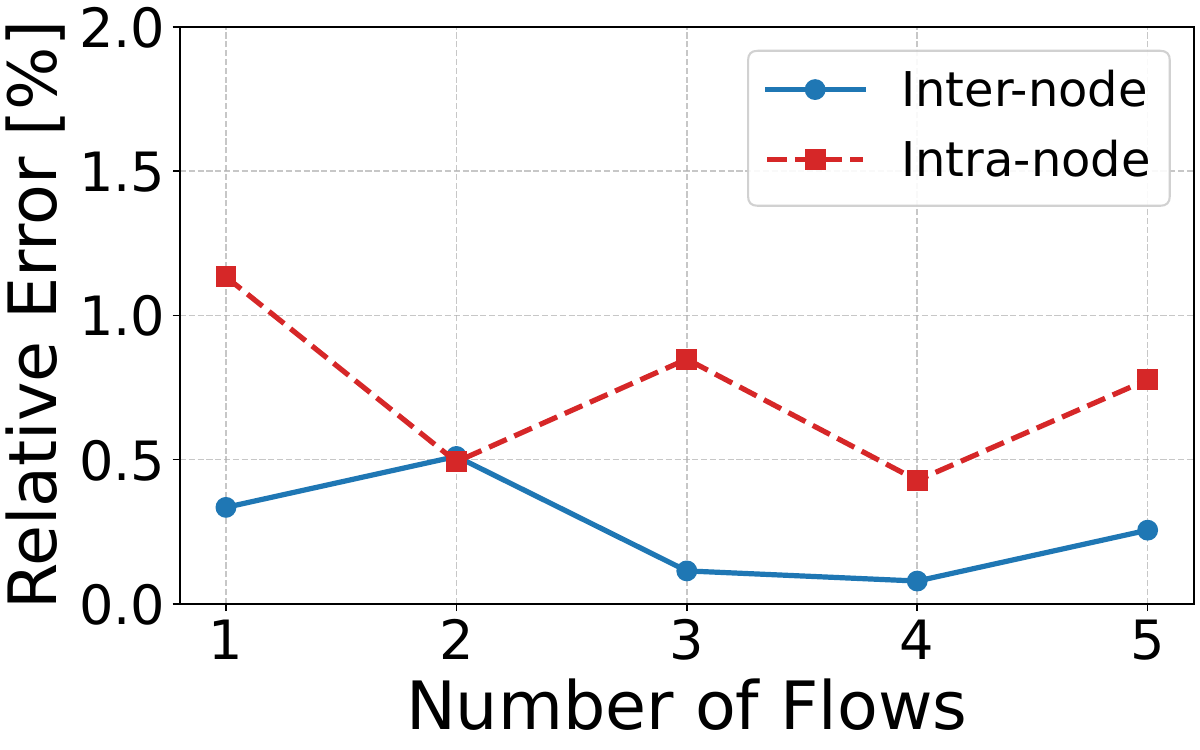}
        \caption{Measurement error.}
        \label{fig:bandwidth_error}
    \end{subfigure}
    \caption{Performance of the measurement module prototype.} 
    \label{fig:bandwidth_analysis}
\end{minipage}
\hfill
\begin{minipage}[b]{0.41\linewidth}
    \captionsetup{type=table}
    \vspace{0pt}
    \centering
    \tiny
    \setlength{\tabcolsep}{2.2pt} 
    \renewcommand{\arraystretch}{2.1}
    \begin{tabular}{|l|ccc|}
        \hline
        \rowcolor{gray!10} \textbf{Component} & \textbf{$M_{acc}$} & \textbf{$M_{cache}$} & \textbf{$M_{buf}$} \\
        \hline
        \textbf{Location} & \makecell{GPU \\ Global} & \makecell{GPU \\ Shared} & \makecell{Host \\ DRAM} \\
        \hline
        \rowcolor{gray!10} \textbf{Size Formula} & \makecell{$S_{acc} \times$ \\ $(NGPU+1)$} & \makecell{$S_{cache} \times$ \\ $TB$} & \makecell{$S_{buf} \times$ \\ $TB \times SRV$} \\
        \hline
        \makecell{\textbf{Size} \\ NGPU=128\\ SRV=8, \\TB=1024}& 4KB & 24KB & 256KB \\
        \hline
    \end{tabular}
    \caption{\gls{mccl} Memory Overhead.}
    \label{tab:memory_overhead}
\end{minipage}
\hfill
\end{figure}

\subsubsection{Measurement Module}

We first evaluate the feasibility and efficiency of the \gls{mccl} measurement module. Our evaluation focuses on two sets of metrics: (1) \textit{throughput} and \textit{memory overhead}; and (2) \textit{measurement accuracy}. To quantify these metrics, we design targeted benchmarks.


\smallskip
\noindent
\textbf{Overhead.} We evaluate \textit{throughput overhead} by comparing \gls{mccl} direct point-to-point throughput with and without our module, varying the \texttt{reduceCopy} chunk size. This setup represents a worst-case stress test: data is immediately available, rendering \texttt{waitPeer} latency zero and fully exposing measurement costs.
\Cref{fig:bandwidth_overhead} illustrates the effective throughput for both inter-node and intra-node transfers as we vary the number of channels, demonstrating that inter-node overhead is negligible because the CPU Proxy manages telemetry without contending for GPU resources. Intra-node overhead peaks at ${\approx}\,\textbf{\SI{13}{\percent}}$ for low channel counts (1--2) and small chunks (\SI{128}{\kilo\byte}) due to granular synchronization, i.e., \texttt{postPeerFence}, and atomic operations. However, this drops to ${\approx}\,\textbf{\SI{2}{\percent}}$ at NVLink saturation (16 channels) and to ${<}\,\textbf{\SI{1}{\percent}}$ with 32 channels (default in H200), confirming the module is non-intrusive for its target operational point.


To preserve capacity for model parameters, \gls{mccl} minimizes \textit{memory overhead} using the three-tiered hierarchy from \Cref{sec:intra-node-measurements}. The total memory footprint comprises three distinct data structures: the Accumulator in GPU Global Memory ($M_{acc}$), the Cache in GPU Shared Memory ($M_{cache}$), and the Buffer in Host-Mapped Memory ($M_{buf}$). The single-entry sizes for these structures are $S_{acc}=\SI{32}{\byte}$, $S_{cache}=\SI{24}{\byte}$, and $S_{buf}=\SI{32}{\byte}$, respectively. $M_{acc}$ scales with communicating peers, while $M_{cache}$ and $M_{buf}$ scale with thread blocks. As shown in \Cref{tab:memory_overhead}, even with 128 GPUs and 1024 TBs, the aggregate memory overhead remains negligible, particularly given the low concurrency of collectives during training.

\smallskip
\noindent
\textbf{Accuracy.}
%
For the \textit{accuracy} assessment, we compare the telemetry obtained from \gls{mccl} against application-level ground truth (pytorch) by varying the number of concurrent flows to emulate network contention. \Cref{fig:bandwidth_error} shows the relative error between the MCCL-reported throughput and the actual available bandwidth. The system demonstrates low estimation error for both inter and intra-node communication, maintaining a consistently low error rate (${<}\,\textbf{\SI{1.2}{\percent}}$). This accuracy allows the Controller to distinguish between healthy connections and gray failures, validating the module's suitability for driving the online optimization loop.

\subsubsection{Hot-Swap Mechanism}
\label{subsubsec:hotswap}

We implement the hot-swap mechanism building on top of MSCCL for on-demand connection and dynamic algorithm loading~\cite{msccl}. 
When the controller issues a new schedule version, the system delegates the transition to a dedicated background thread. This is responsible for asynchronously loading the new algorithmic routing and establishing the necessary peer-to-peer connections without interrupting ongoing communications. If any, existing connections are reused, matching NCCL's default behavior~\cite{nccl-runtimeconnect}.

Once setup completes, the background thread briefly pauses the  issuance of new collectives on the communicator and executes an out-of-band (OOB) \textsc{AllGather}, adapted from MCCS~\cite{Yongji2024mccs}. Each GPU reports the latest version among its successfully loaded schedules and the sequence number of its last-issued collective operation. 
The minimum schedule version ensures that the selected schedule is available on every GPU, while the maximum sequence number $K$ defines the schedule transition point: all operations up to $K$ use the old schedule, and subsequent collective operations use the new one. Since NCCL preserves operation order within a communicator and serializes their device execution~\cite{ncclgroups,nccl-enqueue}, the two schedules never execute concurrently, and no draining of in-flight operations is required. Swaps are scoped independently to each communicator and do not interfere with one another.


\begin{figure}[!t]
\begin{minipage}[b]{0.325\linewidth}
    \captionsetup{type=table}
    \centering
    \tiny
    \setlength{\tabcolsep}{2.6pt}
    \renewcommand{\arraystretch}{1.34} 
    \begin{tabular}[t]{|l|l|}
        \hline
        \rowcolor{gray!10}
        \textbf{Servers}               & 2 hosts $\times$ 4 GPUs \\
        \hline
        \textbf{GPUs}                  & NVIDIA H200 \\
        \hline
        \rowcolor{gray!10}
        \textbf{Intra-node link}       & \begin{tabular}{@{}l@{}}NVLink~4.0 (NVSwitch)\\NV18, $\sim$900\,GB/s/GPU\end{tabular} \\
        \hline
        \textbf{Inter-node NIC}        & \begin{tabular}{@{}l@{}}ConnectX-7\\1 NIC/GPU, PCIe~5.0 $\times16$\end{tabular} \\
        \hline
        \rowcolor{gray!10}
        \textbf{Inter-node link rate}        & 200\,Gbps IB \\
        \hline
        \textbf{Out-of-band NIC}       & Mellanox ConnectX-5 \\
        \hline
        \rowcolor{gray!10}
        \textbf{Out-of-band link rate} & 100\,Gbps Ethernet \\
        \hline
    \end{tabular}
    \caption{Prototype testbed.}
    \label{tab:prototype}
\end{minipage}
\hfill
\begin{minipage}[b]{0.66\linewidth}
    \captionsetup{type=figure}
    \vspace{-2.0pt}
    \hfill
    \centering
    \begin{subfigure}[b]{0.491\linewidth}
        \includegraphics[width=\linewidth, valign=t]{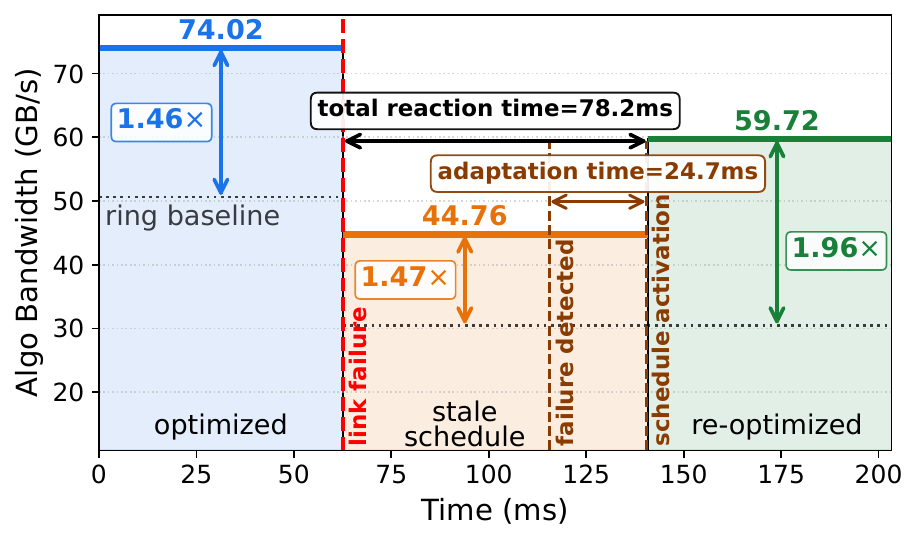}
        \caption{Algo BW before/after failure.}
        \label{subfig:proto_bw_timeline}
    \end{subfigure}
    \hfill
    \begin{subfigure}[b]{0.491\linewidth}
        \includegraphics[width=\linewidth, valign=t]{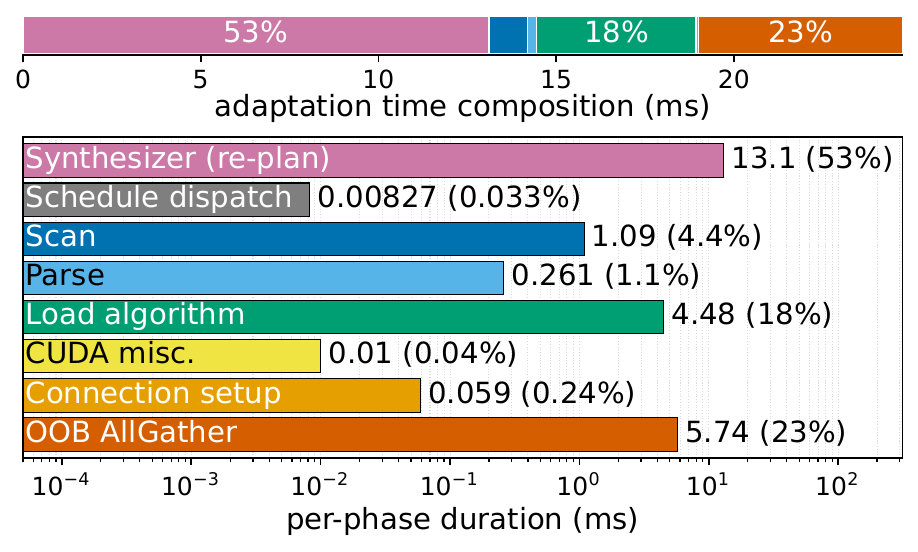}
        \caption{Breakdown of adaptation time}
        \label{subfig:proto_swap_decomposition}
    \end{subfigure}
    \hfill
    \caption{Prototype evaluation of MCCL.}
    \label{fig:prototype}
\end{minipage}
\end{figure}

\Cref{subfig:proto_bw_timeline} illustrates the reaction of the system during an inter-node link failure, which halves the available bandwidth between GPU ranks 0 and 4 (the first GPU of each node) to \SI{100}{\Gbps}. We evaluate a \SI{16}{\mega\byte} per-GPU \textsc{AllGather}, matching the per-GPU input tensor size of the dominant collective in Llama 405B. Under the stale schedule, the degradation reduces algorithm bandwidth~\cite{algobw} from \SI{74.02}{\giga\byte\per\second} to \SI{44.76}{\giga\byte\per\second}. The controller detects the failure in \SI{53.1}{\milli\second}; this window is primarily spent gathering sufficient degraded telemetry samples to prevent false positives from transient anomalies (we use a conservative smoothing factor $\alpha=0.012$). The controller then runs the synthesizer and dispatches the new schedule to the executors, whose Online Rescheduler activates it \SI{24.7}{\milli\second} after detection. Meanwhile, collectives continue running on the stale schedule: all in-flight operations (here, a single \textsc{AllGather}) complete on it, while the next submitted collective switches to the new schedule, starting execution \textbf{\SI{0.4}{\milli\second}} later for a total reaction time of \textbf{\SI{78.2}{\milli\second}}. Once executed, algorithm bandwidth recovers to \SI{59.72}{\giga\byte\per\second} without interrupting training execution.

\Cref{subfig:proto_swap_decomposition} breaks down the adaptation time of \textbf{\SI{24.7}{\milli\second}}: \textbf{\SI{13.1}{\milli\second}} is spent in the synthesizer invoked by the Schedule Optimizer, while the hot-swap takes \textbf{\SI{11.6}{\milli\second}}, mainly for OOB synchronization (\textbf{\SI{5.74}{\milli\second}}) and algorithm loading (\textbf{\SI{4.48}{\milli\second}}). As node count grows, re-optimization increasingly dominates this latency (cf. \Cref{subsec:algorithm-evaluation}), making the swap overhead comparatively small. Moreover, hot-swap operations run on dedicated threads, remaining off the critical execution path; only the OOB \textsc{AllGather} pauses the issuing of new collectives for \textbf{\SI{5.74}{\milli\second}}. In practice, for pipelined large-scale training workloads this brief CPU-side pause overlaps with ongoing GPU execution.

\begin{table}[t]
\tiny
\begin{center}
\setlength{\tabcolsep}{2.6pt}
\begin{tabular}{|l|cc|cc|c|ccc|cc|cc|}
\hline
\multirow{2}{*}{\textbf{Metric}} & \multicolumn{5}{c|}{\textbf{Llama 405B}} & \multicolumn{7}{c|}{\textbf{DeepSeek 671B}} \\
\cline{2-13}
 & \multicolumn{2}{c|}{\textbf{AllGather}} & \multicolumn{2}{c|}{\textbf{ReduceSc.}} & \multicolumn{1}{c|}{\textbf{AllReduce}} & 
 \multicolumn{3}{c|}{\textbf{AllGather}} & \multicolumn{2}{c|}{\textbf{ReduceSc.}} & \multicolumn{2}{c|}{\textbf{AllToAll}} \\
\hline
\textbf{Group (\#GPUs)} & DP (64) & TP (16) & DP (64) & TP (16) & TP (16)
& DP (1024) & DP\_EP (16) & EP (64) & DP (1024) & DP\_EP (16) & EP (64) & EP (64) \\
\textbf{Size} & 2.6 GB & 256 MB & 5.3 GB & 256 MB & 256 MB 
& 1.2 GB & 0.9 GB & 32 KB & 2.4 GB & 1.9 GB & 448 MB & 231 MB \\
\textbf{Frequency} & 16 & 524K & 16 & 524K & 16K 
& 1 & 64 & 768 & 1 & 64 & 2304 & 768 \\
\textbf{Training time} & 0.167\% & 18.0\% & 0.555\% & 15.6\% & 7.36e-4\% 
& 0.136\% & 4.11\% & 0.141\% & 0.266\% & 8.22\% & 44.5\% & 7.66\% \\
\hline
\end{tabular}
\end{center}
\caption{Llama 405B and DeepSeek 671B (batch size 16K) collectives for one pipeline stage with 1024 GPUs.}
\label{tab:workloads}
\end{table}

\begin{figure}[!t]
    \centering
    \begin{subfigure}[b]{0.245\linewidth}
        \includegraphics[width=\linewidth]{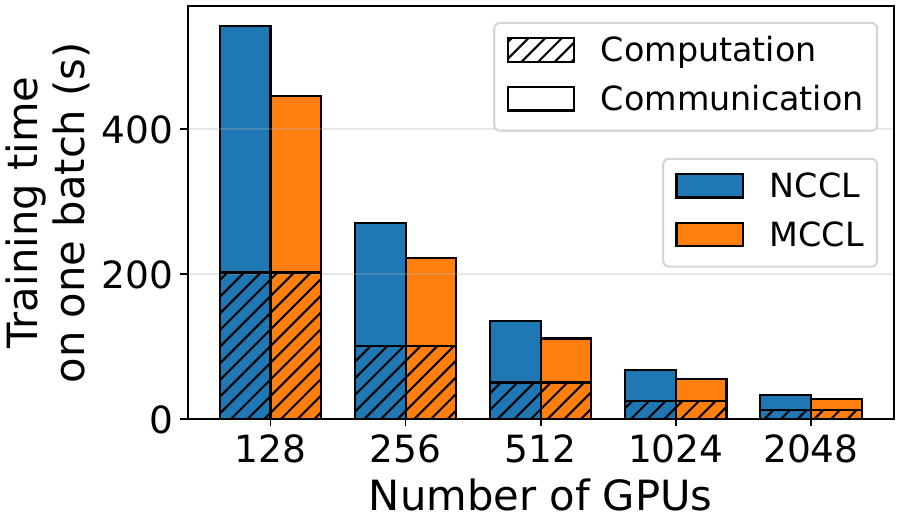}
        \caption{Llama405B, Fail 1.}
        \label{subfig:end-to-end-llama-1}
    \end{subfigure}
    \hfill
        \begin{subfigure}[b]{0.245\linewidth}
        \includegraphics[width=\linewidth]{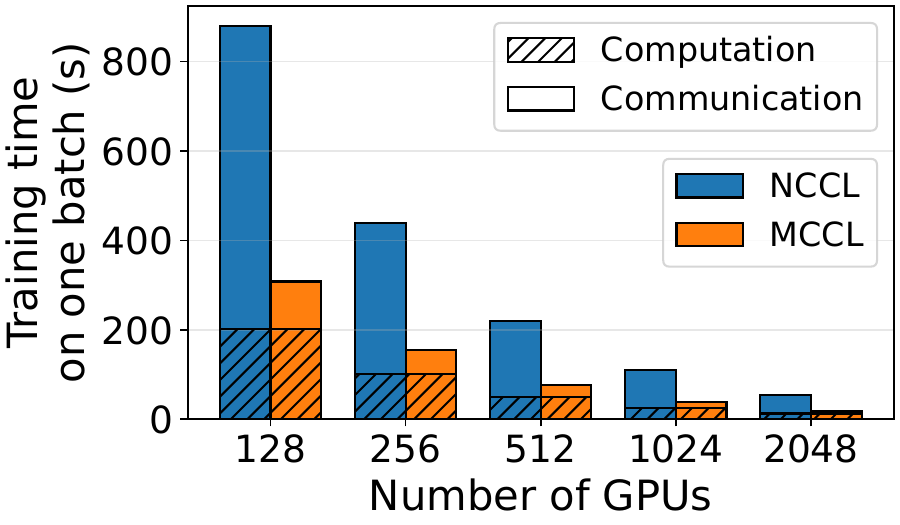}
        \caption{Llama405B, Fail 2.}
        \label{subfig:end-to-end-llama-2}
    \end{subfigure}
    \hfill
        \begin{subfigure}[b]{0.245\linewidth}
        \includegraphics[width=\linewidth]{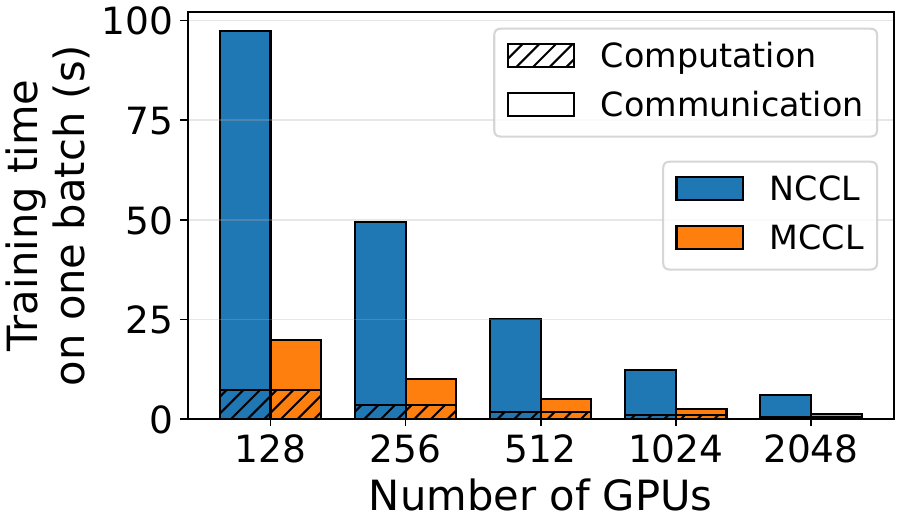}
        \caption{DeepSeek671B, Fail 2.}
        \label{subfig:end-to-end-deepseek-2}
    \end{subfigure}
    \hfill
    \begin{subfigure}[b]{0.245\linewidth}
        \includegraphics[width=\linewidth]{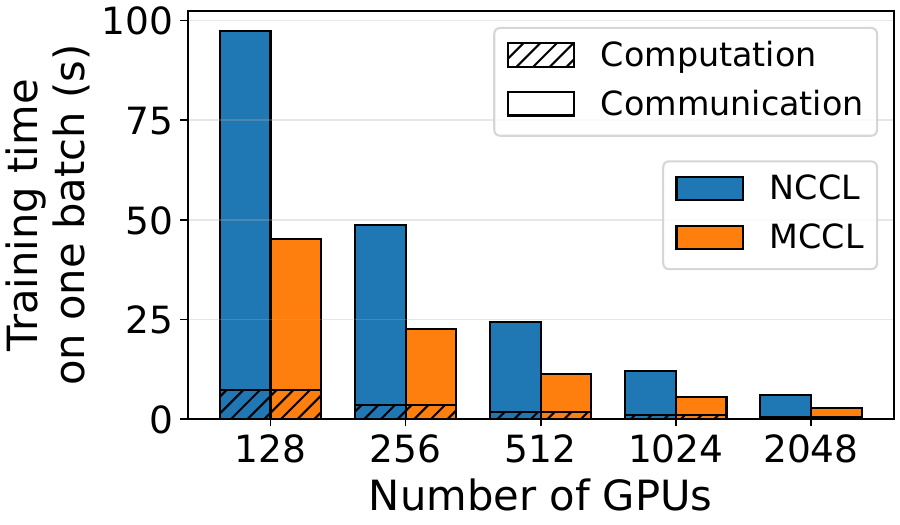}
        \caption{DeepSeek671B, Fail 3.}
    \label{subfig:end-to-end-deepseek-3}
    \end{subfigure}
     \caption{Training time for Llama405B (a),(b) and DeepSeek671B(c),(d) with different failures profiles.}
    \label{fig:end-to-end}
\end{figure}

\subsection{End-to-End Validation}
\label{subsec:end-to-endValidation}

To validate the end-to-end benefits of our approach, we run large-scale simulations in SimAI. We use an HPN topology with segments of 128 GPUs each (16 hosts) and scale the cluster from 128 to 2048 GPUs; we apply the same three failure scenarios as in \Cref{subsec:algorithm-evaluation}. Because detection parameters ($\Gamma, \alpha$) are inherently cluster-specific, calibrating them to our simulation offers little generalizable insight. We therefore decouple condition detection from schedule adaptation, evaluating \gls{mccl}'s re-synthesis and failover in isolation under prompt failure notifications, in regimes where adaptation proceeds without schedule thrashing. As a baseline, we compare against plain NCCL. We exclude TECCL due to impractical runtime solving time, and ForestColl as it cannot optimize \textsc{AllToAll} collectives, while it delivers the same schedule quality as \gls{mccl} for \textsc{AllGather} collectives.

We design two sets of workloads that mimic the training of Llama3.1 with 405B parameters and DeepSeek-v3 with 671B parameters and 256 experts~\cite{grattafiori2024llama3,liu2024deepseek}.
We configure a pipeline parallelism of 16 and only simulate one pipeline stage as traffic between stages is point-to-point. Therefore, 
our simulations with 128--2048 GPUs correspond to training the full model on 2048--32768 GPUs. We execute a forward and backward pass on a batch of 16K tokens.
\Cref{tab:workloads} details collective communications of both workloads, each with its share over the total training time. In our settings, collective communications account for ${\approx}\,\textbf{\SI{34}{\percent}}$ of Llama's training time, and for ${\approx}\,\textbf{\SI{65}{\percent}}$ of DeepSeek's.%
\footnote{In our end-to-end experiments, we do not optimize the 1024-GPU \textsc{AllGather} in DeepSeek, as it accounts for only $\textbf{\SI{0.136}{\percent}}$ of the total execution time, making the associated synthesizer overhead unjustified in practical settings. Moreover, \textsc{ReduceScatter} collectives operate at the host level, as the simulator does not support in-network reduction.}

\Cref{fig:end-to-end} reports the training time for Llama and DeepSeek under the failures described above. The results of all failure profiles are reported in \Cref{app:endtoend}. 
When no failures are applied, MCCL provides no advantages over NCCL (cf. \Cref{app:endtoend}); this is because this topology has, ring-wise, a perfect ratio between inter- and intra-node bandwidth. 
Under failure conditions, MCCL significantly reduces training time, with peaks of $\mathbf{2.85\times}$ and $\mathbf{4.99\times}$ under failure~2 for Llama and DeepSeek, respectively. Finally, MCCL scales effectively with increasing GPU counts, yielding similar relative gains across different cluster sizes.
\section{Related Work}
\label{sec:rw}

\smallskip\noindent
\textbf{Offline schedule synthesis.}
A large body of work synthesizes collective schedules offline. TACCL~\cite{shah2023taccl}, TE-CCL~\cite{xuting2024teccl}, and ForestColl~\cite{zhao2025forestcoll} build topology-aware schedules from communication sketches, traffic-engineering formulations, and spanning trees, respectively, outperforming vendor-provided collectives.
However, they rely on static, offline profiling of the topology at peak capacity and cannot promptly react to link degradation, multi-tenant interference, or gray failures.
The concurrent work OptCCL~\cite{shapley2026optccl} synthesizes provably optimal schedules from a multi-commodity flow \gls{lp} solved via Dantzig--Wolfe decomposition, with an \textsc{AllToAll} formulation close to ours (\Cref{subsec:a2a_optimization}).
OptCCL scales the decomposition by exploiting topology symmetry, which gray failures inherently break, forcing a fallback to a cold start.
\gls{mccl} instead exploits the locality of re-optimization, reducing re-synthesis time without relying on symmetry.
The two approaches are complementary; moreover, OptCCL neither detects degradations nor redeploys schedules at runtime.

\smallskip\noindent
\textbf{Runtime adaptation.}
A second line of work adapts at runtime, but along axes orthogonal to ours.
FAST~\cite{lei2026fastefficientscheduleralltoall} schedules MoE's skewed, dynamic \texttt{AllToAllv} online, rebalancing over the fast intra-server fabric so that every NIC sends equal volume across the slow inter-server links.
Its adaptation tracks \emph{workload} skew, not \emph{network} state: it models the cluster as two uniform, equal-capacity bandwidth tiers, so its balancing breaks under link degradation, asymmetric capacity, or gray failures.
At the inter-job level, Crux~\cite{cao2024crux} and MCCS~\cite{Yongji2024mccs} arbitrate the network shared by co-located jobs: Crux prioritizes flows by GPU intensity (compute-to-communication ratio), routing GPU-heavy jobs away from contended links, while MCCS, a provider-managed service, reorders ring ranks and pins flows to paths to preserve fairness and reduce congestion.
Both re-optimize only at job arrival/exit or on external triggers, leaving collective schedules unchanged and thus exposed to gray failures.
The concurrent work Theseus~\cite{ding2026theseus} hot-swaps schedules at runtime according to user policies, but neither synthesizes schedules nor measures the network: it relies on slow \gls{milp}-based synthesizers or precomputed candidates, and on NIC statistics that miss intra-host degradations (\eg{} NVLink, PCIe).
\gls{mccl} instead measures each GPU-to-GPU path and re-synthesizes within tens of seconds, only when the bottleneck cut loses saturation.

\smallskip\noindent
\textbf{Execution backends.}
A final line of work provides the execution backends. NCCL and RCCL ship static, built-in algorithms, while MSCCL~\cite{cowan2023msccl}, MSCCL++~\cite{Hwang_2026_msccl++}, and ResCCL~\cite{liu2025resccl} add DSL-defined schedules, portable low-level primitives, and resource-efficient thread-block scheduling, respectively.
They improve expressiveness, portability, and efficiency, but still execute static, offline-generated schedules and cannot adapt to link variability, congestion, or gray failures.

\section{Conclusion}
\label{sec:conclusion}
This paper introduced a \acrfull{mccl} that shifts the paradigm from ``optimize-once'' to ``continuously-adaptive''. By leveraging a controller-executor architecture, \gls{mccl} integrates lightweight, passive, high-fidelity GPU telemetry directly into the optimization loop without disrupting the training workload. 
Our evaluation demonstrates that \gls{mccl} re-optimization algorithms rapidly synthesize schedules adapted to detected bottlenecks, improving solving times 
by up to $\mathbf{4.40{\times}}$ for \textsc{AllGather} and by up to $\mathbf{17.8{\times}}$ for \textsc{AllToAll} compared to static synthesizers. The \gls{mccl} prototype demonstrates the feasibility of the complete closed loop, with low measurement overhead (${<}\,\textbf{\SI{2}{\percent}}$), high accuracy (${<}\,\textbf{\SI{1.2}{\percent}}$ error), and fast schedule swaps (${\approx}\,\textbf{\SI{12}{\milli\second}}$). Finally, large-scale simulations of LLM training workloads show that \gls{mccl} reduces training time by up to $\mathbf{4.99\times}$ under gray failures compared to static schedules.




\smallskip
\noindent\textbf{This work does not raise any ethical issues.}

\newpage

\bibliographystyle{plain}
\bibliography{biblio}

\clearpage
\begingroup
\appendix
\crefalias{section}{appendix}
\crefalias{subsection}{subappendix}
\crefalias{subsubsection}{subsubappendix}
\section{Rail-optimized Topology}
\label{app:topology}

Figure~\ref{fig:railtopology} details the rail-optimized topology used in our evaluation, specifically highlighting a single segment consisting of dual servers with an 8-GPU configuration.


\section{Measurement Matrix}
\label{app:measurements}

Figure~\ref{fig:matrixmeasure} visualizes the passive measurement data structure described in Section~\ref{sec:measurements}. Each GPU fills its corresponding row using \textit{RDMA writes} during collective execution. The diagonal represents high-bandwidth intra-node communication (NVLink), while off-diagonal blocks represent inter-node communication, providing a real-time heatmap of cluster network health.
%

\begin{figure}[h]
\centering
    \begin{minipage}[b]{0.43\textwidth}
    \centering
        \includegraphics[width=0.8\linewidth]{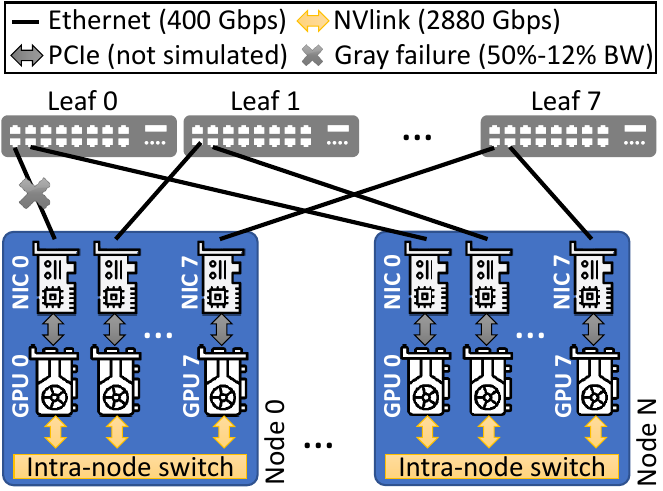} 
        \caption{Rail Optimized topology showing links between GPU nodes and leaf switches.}
    \label{fig:railtopology}
    \end{minipage}\hfill
    \begin{minipage}[b]{0.54\textwidth}
    \centering
        \includegraphics[width=0.74\linewidth]{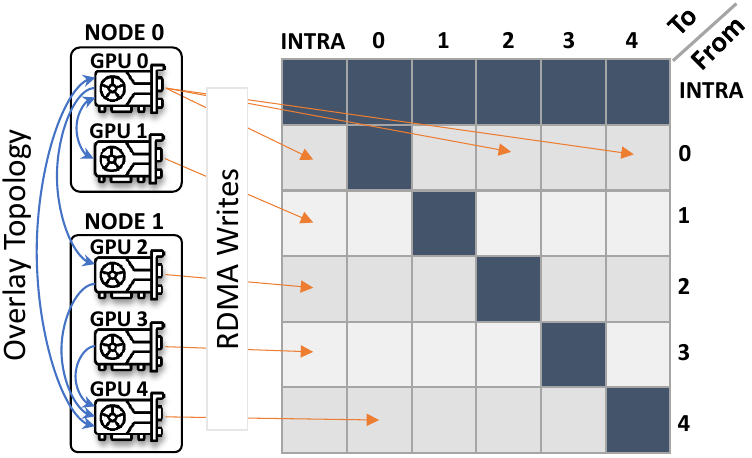} 
        \caption{Per collective type, size, and communicator group point-to-point measurements. Darker blocks indicate intra-node bandwidth, lighter blocks inter-node connections.}
        \label{fig:matrixmeasure}
    \end{minipage}
\end{figure}

\section{Latency-Aware Switch Removal for AllGather}
\label{app:latency_switch_removal}

ForestColl switch removal~\cite{zhao2025forestcoll} only optimizes throughput, ignoring topological diameter and often yielding sparse structures (e.g., rings). Thus the tree-generation phase often builds excessively deep trees. We propose a formulation that bounds tree height and reduces communication steps.

\noindent
\textit{Overlay to Logical Topology ---}
Given a node with $n$ GPUs and one high-speed switch, and being $G_i$ the
subgraph of the $i$-th node's GPUs, switch, and links, switch removal
replaces $G_i$ with a logical topology $\bar G_i=(\bar V_i,\bar A_i)$ where
vertices $\bar V_i$ are the node's GPUs and arcs $\bar A_i$ connect all pairs of GPUs linked by a communication path.
Each GPU's outgoing capacity is split across its logical links.

\noindent
\textit{Improved Switch Removal Formulation ---}
We remove one switch at a time. When processing switch $w$, $\bar G_i{=}(\bar V_i,\bar A_i)$ is the local graph on the predecessors
$\delta^-(w)$ and successors $\delta^+(w)$ of $w$; its candidate \emph{logical
links} $\bar A_i{=}\{(u,v)\,:\,u{\in}\delta^-(w),\,v{\in}\delta^+(w),\,u{\neq}v\}$ join each
predecessor to each successor. Each $a=(u,v)$ carries a capacity variable
$y_a\ge 0$, and we seek values that reproduce, between predecessors and
successors, the connectivity $w$ provided. We write $N=|\bar V_i|$, $c_a$ for
physical capacities, and $c_{uw}$ (resp. $c_{wv}$) for the ingress (resp. egress)
capacity of $w$.

To keep trees shallow, a secondary objective spreads each node's traffic over
many successors. Being $\delta^+(w)$ the set of successors of $w$, we add a convex, piecewise-linear penalty $h_{uv}$ whose $k$-th
piece has slope $k$ and whose breakpoints are integer multiples of the width
\begin{smallmath}
\setlength{\abovedisplayskip}{2pt}
\setlength{\belowdisplayskip}{2pt}
  w_{uv}=\frac{c_{uw}}{|\delta^+(w)|-1};
\end{smallmath}%
%
thus $w_{uv}$ only depends on
$u$ and equals the fair per-successor share of $u$'s ingress---we round $w_{uv}$ down to the nearest integer to keep the constraint coefficients integral. Since the marginal penalty grows with
the capacity placed on a single link, concentrating flow costs more than
spreading it, biasing the logical topology toward a low-depth star.
The throughput objective takes priority, so we combine the two
lexicographically with a single big-$M$ weight: any $M$ strictly larger than the
penalty's maximum value suffices. At optimality, each $h_{uv}$ equals
$\max_{1\le k\le K}\{k\,y_a{-}w_{uv}\tfrac{k(k-1)}{2}\}{\le}K\,y_a$, with
$K{=}|\delta^+(w)|{-}1{\le}N{-}1$. Summing over $\bar A_i$ and using the predecessor
bounds $\sum_{v}y_{uv}\le c_{uw}$,

\begin{smallmath}
\setlength{\abovedisplayskip}{3pt}
\setlength{\belowdisplayskip}{3pt}
  \sum_{a=(u,v)\in\bar A_i} h_{uv}
  \;\le\; K\!\!\sum_{a\in\bar A_i}\!\! y_a
  \;\le\; (N-1)\!\!\sum_{u\in\delta^-(w)}\!\! c_{uw}
  \;\le\; (N-1)\,N\,c_{\max}
  \;<\; N^2 c_{\max},
\end{smallmath}%
where $c_{\max}=\max_a c_a$. We therefore set $M=N^2 c_{\max}$, so improving
throughput by one unit always outweighs any attainable change in the penalty; the
penalty only discriminates among throughput-optimal solutions.
The switch removal is formulated as follows:
%

\vspace{-5pt}
\noindent
\begin{minipage}{0.58\textwidth}
\begin{smallalign}
  \setlength{\abovedisplayskip}{0pt}
  \setlength{\belowdisplayskip}{3pt}
  &\textstyle \max \ M \sum_{a\in\bar A_i} y_a - \sum_{a=(u,v)\in\bar A_i} h_{uv} \label{partial:objective_sw}\\
  &\textstyle h_{uv} \ge k\,y_a - w_{uv}\,\frac{k(k-1)}{2} \quad \forall a=(u,v)\in\bar A_i,\ k\in\{1,\dots,K\} \label{partial:piecewise}\\
  &\textstyle \sum_{v:\,(u,v)\in\bar A_i} y_{uv} \le c_{uw} \quad \forall u\in\delta^-(w), \label{partial:cap_pred}
\end{smallalign}
\end{minipage}%
\hfill
\begin{minipage}{0.41\textwidth}
\begin{smallalign}
  \setlength{\abovedisplayskip}{0pt}
  \setlength{\belowdisplayskip}{3pt}
  &\textstyle \sum_{u:\,(u,v)\in\bar A_i} y_{uv} \le c_{wv} \quad \forall v\in\delta^+(w), \label{partial:cap_succ}\\
  &\textstyle \sum_{a\in A_{\bar V'}} y_a \ge MT - OF_{V'} \quad \forall V'\subseteq V, \label{partial:mincut}\\
  &\textstyle y_a,\ h_{uv} \ge 0 \quad \forall a=(u,v)\in\bar A_i. \label{partial:bound}
\end{smallalign}
\end{minipage}
\vspace{7pt}


In~\eqref{partial:objective_sw}, $M\sum_a y_a$ maximizes the total logical
capacity installed between the predecessors and successors of $w$ (primary
throughput goal) and $-\sum h_{uv}$ applies the star-inducing penalty (secondary
goal). Constraints~\eqref{partial:piecewise} define $h_{uv}$ as the lower envelope
of the convex penalty above. Constraints~\eqref{partial:cap_pred}--%
\eqref{partial:cap_succ} preserve capacity: links leaving a predecessor $u$ carry
at most its ingress capacity $c_{uw}$, and links entering a successor $v$ at most
its egress capacity $c_{wv}$, so the logical topology is realizable on the fabric.
Constraints~\eqref{partial:mincut} are the throughput (min-cut) constraints:
across every cut $V'\subseteq V$, the flow must be at least the maximum throughput
$MT$, where $OF_{V'}$ is the flow crossing outside the switch's logical links and
$A_{\bar V'}=A_{V'}\cap\bar A_i$ is the logical part of the cut.
Finally,~\eqref{partial:bound} enforces non-negativity.

The min-cut constraints are exponential in number, so we solve the model by
cutting planes: at each iteration we compute, for every destination GPU, the
maximum flow from the source under the current $y$; if any is below $MT$ we add
the corresponding cut~\eqref{partial:mincut}, otherwise the relaxation is optimal.
By cutting-plane theory~\cite{Grotscheletal1981}, the number of generated
constraints is polynomial, $O(n^\kappa)$, in the number of variables $n$
($\kappa$ constant).

\Cref{fig:FCPhyLog} shows an example intra-node overlay-to-logical mapping, and
\Cref{fig:FCCuttingPlane} illustrates a cut added by the cutting-plane algorithm
during switch removal, ensuring the generated trees respect the bandwidth limits
identified by the max-flow formulation.

\begin{figure}[h]
\centering
    \begin{minipage}[b]{0.35\textwidth}
    \centering
        \includegraphics[width=0.75\linewidth]{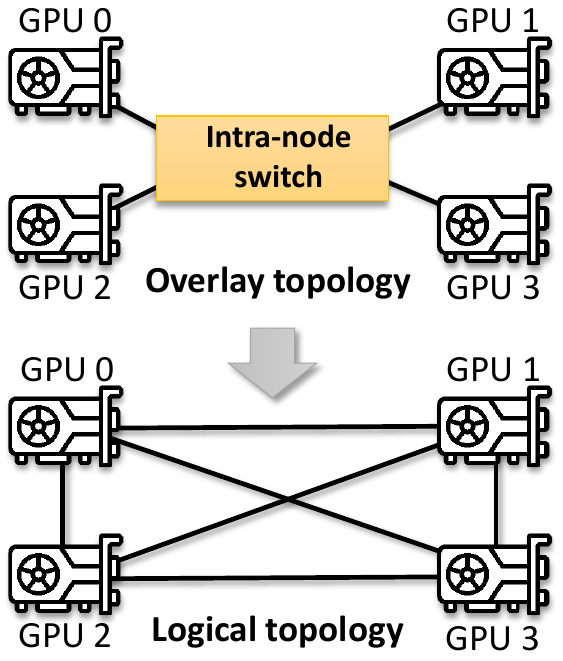} 
        \caption{Overlay to Logical topology.}
    \label{fig:FCPhyLog}
    \end{minipage}\hfill
    \begin{minipage}[b]{0.61\textwidth}
    \centering
        \includegraphics[width=0.82\linewidth]{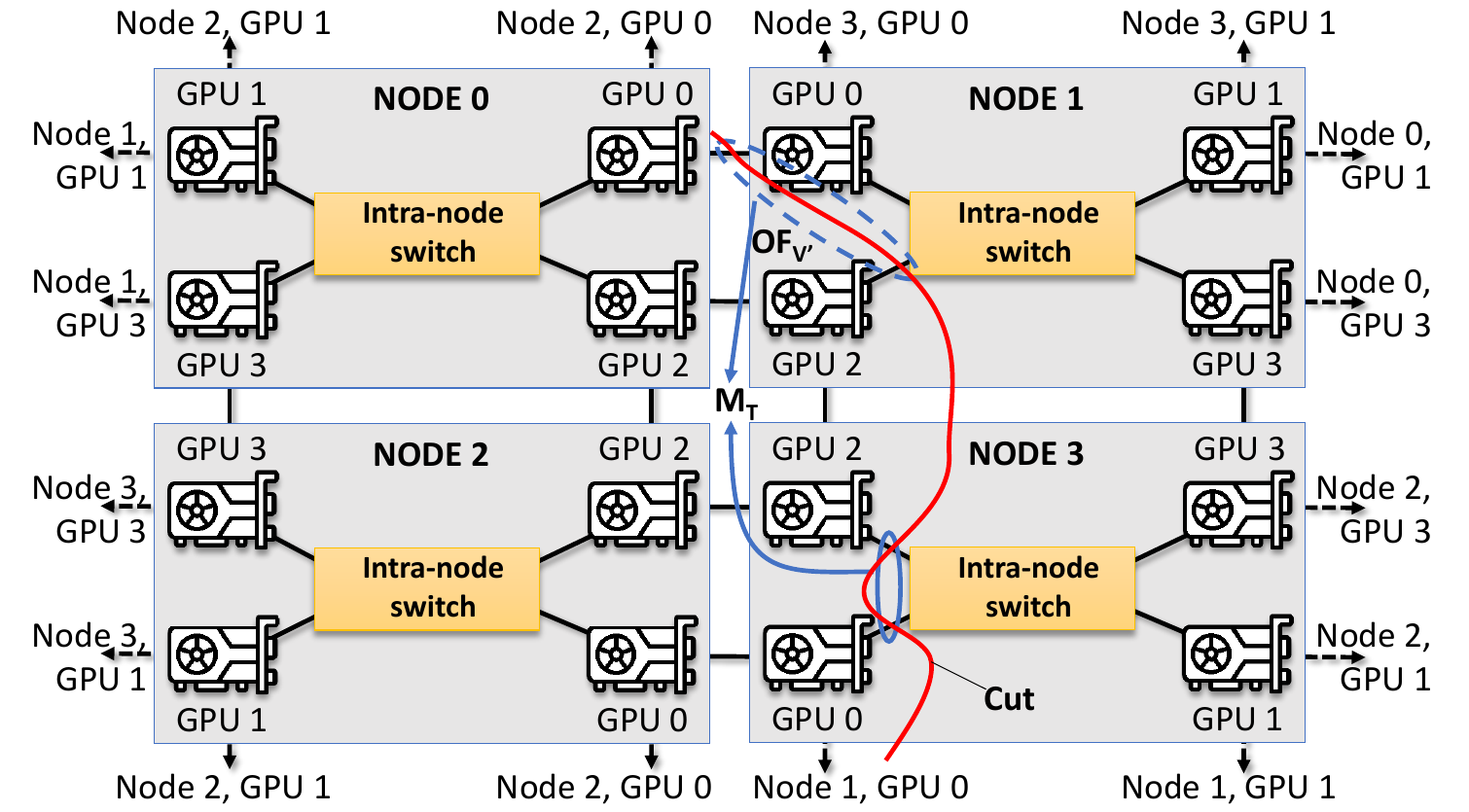} 
        \caption{Example cut induced by the max-flow in the switch removal phase, enforcing capacity constraints on the logical topology.}
        \label{fig:FCCuttingPlane}
    \end{minipage}
\end{figure}

\section{Minimizing AllToAll Steps via Binary Search}
\label{app:a2a_minsteps}
The online loop fixes the step bound to $m_s = |V|-1$ for fast re-synthesis under persistent anomalies, but when computational time allows, it's possible to minimize steps without sacrificing throughput. Let $MT$ be the optimal throughput at $m_s = |V|-1$. We binary-search the smallest bound $m_s^* \le |V|-1$ that still attains $MT$: from $L=1$, $U=|V|-1$, each iteration reduces the time-expanded graph $\tilde{G}$ to $m_{test} = \lfloor (L+U)/2 \rfloor$ steps and runs CG. If $MT$ is reached we set $U = m_{test}-1$ and cache the schedule; if the model is infeasible or falls short of $MT$ we set $L = m_{test}+1$. To accelerate convergence we warm-start via MCCL's mechanism, injecting columns from prior iterations into the RMP. The search converges logarithmically to the minimal throughput-optimal step count.

\section{End-to-End Validation}
\label{app:endtoend}

\Cref{fig:end-to-end-llama-appendix,fig:end-to-end-deepseek-appendix} show complete end-to-end training time for Llama 405B and DeepSeek 671B, expanding on the results from Section~\ref{subsec:end-to-endValidation} under all failure scenarios: no failure, Failure 1 (Switch), Failure 2 (Link degradation), and Failure 3 (GPU replacement).

\smallskip
\begin{figure}[h]
    \centering
    \begin{subfigure}[b]{0.24\textwidth}
        \includegraphics[width=\linewidth]{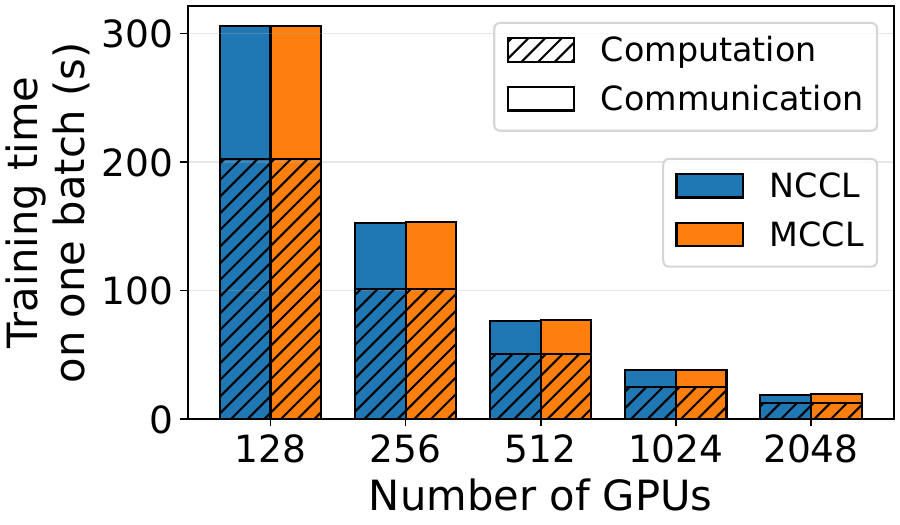}
        \caption{No failure}
        \label{subfig:llama-no-fail}
    \end{subfigure}
    \hfill
    \begin{subfigure}[b]{0.24\textwidth}
        \includegraphics[width=\linewidth]{plots/end_to_end/Llama405B_failure_1.pdf}
        \caption{Fail 1 (Switch)}
        \label{subfig:llama-fail1}
    \end{subfigure}
    \hfill
    \begin{subfigure}[b]{0.24\textwidth}
        \includegraphics[width=\linewidth]{plots/end_to_end/Llama405B_failure_2.pdf}
        \caption{Fail 2 (Link)}
        \label{subfig:llama-fail2}
    \end{subfigure}
    \hfill
    \begin{subfigure}[b]{0.24\textwidth}
        \includegraphics[width=\linewidth]{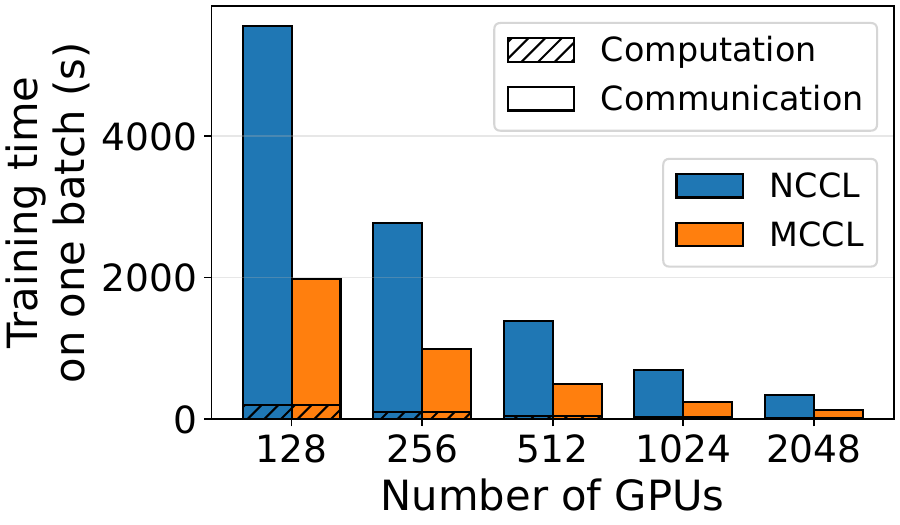}
        \caption{Fail 3 (GPU)}
        \label{subfig:llama-fail3}
    \end{subfigure}
    
    \caption{\textbf{Llama 405B Training Time.} Compare NCCL and MCCL for one pipeline stage (out of 16).}
    \label{fig:end-to-end-llama-appendix}
\end{figure}

\begin{figure}[h]
    \centering
    \begin{subfigure}[b]{0.24\textwidth}
        \includegraphics[width=\linewidth]{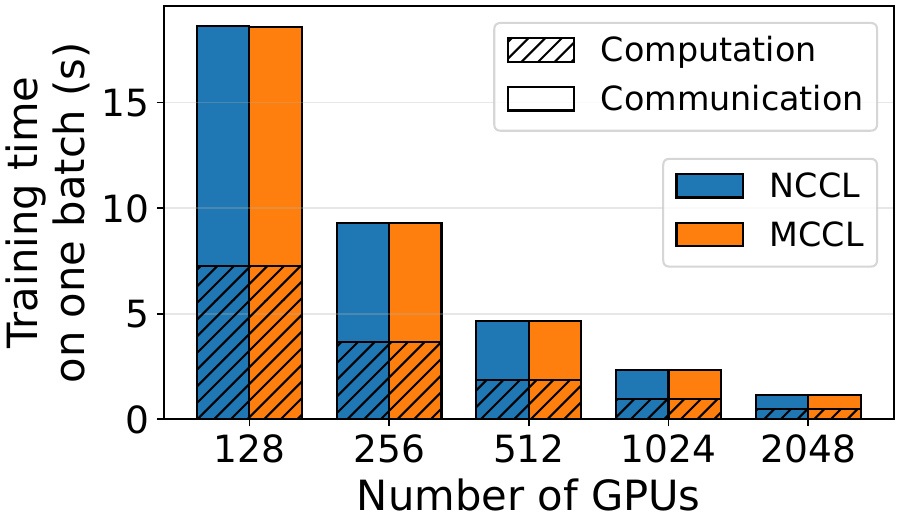}
        \caption{No failure}
        \label{subfig:deepseek-no-fail}
    \end{subfigure}
    \hfill
    \begin{subfigure}[b]{0.24\textwidth}
        \includegraphics[width=\linewidth]{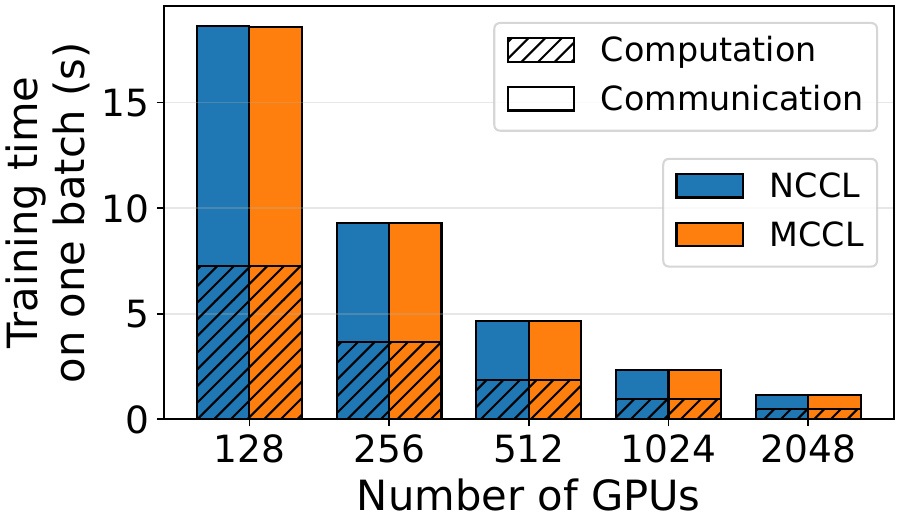}
        \caption{Fail 1 (Switch)}
        \label{subfig:deepseek-fail1}
    \end{subfigure}
    \hfill
    \begin{subfigure}[b]{0.24\textwidth}
        \includegraphics[width=\linewidth]{plots/end_to_end/DeepSeek671B_failure_2.pdf}
        \caption{Fail 2 (Link)}
        \label{subfig:deepseek-fail2}
    \end{subfigure}
    \hfill
    \begin{subfigure}[b]{0.24\textwidth}
        \includegraphics[width=\linewidth]{plots/end_to_end/DeepSeek671B_failure_3.pdf}
        \caption{Fail 3 (GPU)}
        \label{subfig:deepseek-fail3}
    \end{subfigure}
    
    \caption{\textbf{DeepSeek 671B Training Time.} Compare NCCL and MCCL for one pipeline stage (out of 16).}
    \label{fig:end-to-end-deepseek-appendix}
\end{figure}

\section{Rounding}
\label{app:a2a_rounding}
\Cref{tab:rounding} shows results for the rounding process. The rounding error is computed between the optimal \gls{milp} solution and the rounded \gls{lp}. We used 60 cores, 1TB of RAM and a time limit of 1 hour for each \{topology-\#chunks\} combination and the \#chunks is searched between 1 and 10.

\smallskip
\begin{table}[h]
\centering
\footnotesize
\setlength{\tabcolsep}{3pt}
\renewcommand{\arraystretch}{0.8}
\begin{tabular}{|l|c|c|c|c|c|c|c|c|c|c|c|c|}
\hline
\multirow{2}{*}{\textbf{Metric}} & \multicolumn{4}{c|}{\textbf{1x2x8}} & \multicolumn{4}{c|}{\textbf{1x4x8}} & \multicolumn{4}{c|}{\textbf{1x8x8}} \\ \cline{2-13}
& \textbf{Base} & \textbf{Fail 1} & \textbf{Fail 2} & \textbf{Fail 3} & \textbf{Base} & \textbf{Fail 1} & \textbf{Fail 2} & \textbf{Fail 3} & \textbf{Base} & \textbf{Fail 1} & \textbf{Fail 2} & \textbf{Fail 3} \\
\hline
\textbf{LP Bound} & 0.6250 & 0.6250 & 4.4531 & 1.0000 & 16.6250 & 16.6250 & 14.8125 & 4.8065 & 2.3750 & 7.1250 & 6.3438 & 2.3492 \\
\textbf{Rounded Obj} & 0.6250 & 0.6250 & 4.4444 & 1.0000 & 16.6250 & 16.6250 & 14.7778 & 4.8000 & 2.3750 & 7.1250 & 6.3333 & 2.3333 \\
\textbf{MILP Obj} & 0.6250 & 0.6250 & 4.4444 & 1.0000 & 16.6250 & 16.6250 & 14.7778 & 4.8000 & 2.3750 & 7.1250 & 6.3333 & 2.3333 \\
\hline
\textbf{Error (\%)} & 0.00 & 0.00 & 0.00 & 0.00 & 0.00 & 0.00 & 0.00 & 0.00 & 0.00 & 0.00 & 0.00 & 0.00 \\
\hline\hline
\multirow{2}{*}{\textbf{Metric}} & \multicolumn{4}{c|}{\textbf{1x16x8}} & \multicolumn{4}{c|}{\textbf{1x32x8}} & \multicolumn{4}{c|}{\textbf{1x64x8}} \\ \cline{2-13}
& \textbf{Base} & \textbf{Fail 1} & \textbf{Fail 2} & \textbf{Fail 3} & \textbf{Base} & \textbf{Fail 1} & \textbf{Fail 2} & \textbf{Fail 3} & \textbf{Base} & \textbf{Fail 1} & \textbf{Fail 2} & \textbf{Fail 3} \\
\hline
\textbf{LP Bound} & 0.3750 & 3.3750 & 3.0000 & 1.2205 & 1.6250 & 1.6250 & 1.4531 & 0.5608 & 0.1250 & 0.3750 & 0.6719 & 0.1722 \\
\textbf{Rounded Obj} & 0.3750 & 3.3750 & 3.0000 & 1.1915 & 1.6250 & 1.6250 & 1.2500 & 0.5294 & 0.1250 & 0.3750 & 0.5625 & 0.1613 \\
\textbf{MILP Obj} & 0.3750 & 3.3750 & 3.0000 & 1.2174 & 1.6250 & 1.6250 & -* & -* & 0.1250 & 0.3750 & -* & -* \\
\hline
\textbf{Error (\%)} & 0.00 & 0.00 & 0.00 & 2.13 & 0.00 & 0.00 & - & - & 0.00 & 0.00 & - & - \\
\hline
\end{tabular}
\caption{Single-Segment Experiments Error Results (*not all MILP results completed due to out-of-memory).}
\label{tab:rounding}
\end{table}


\section{Compact Model and Derivation of the Extensive \textsc{AllToAll} Formulation}
\label{app:a2a_derivation}

This appendix derives the extensive formulation \eqref{a2a:obj}--\eqref{a2a:bound} by stating a compact, arc-based multi-commodity flow \gls{milp} (\Cref{app:a2a:compact}), relaxing it to a max--min concurrent-flow \gls{lp} (\Cref{app:a2a:relax}), and applying a Dantzig--Wolfe decomposition to recover the master problem and pricing subproblem of \Cref{subsec:a2a_optimization} (\Cref{app:a2a:dw}). We reuse all notation defined in \Cref{subsec:a2a_optimization}.

\vspace{-3pt}
\subsection{Capacities and commodities}
\label{app:a2a:graph}

We extend $\tilde G$ with one \emph{super-sink} $\sigma_d$ per destination GPU
$d\in V$, reached through \emph{exit} arcs $\bigl((d,t),\sigma_d\bigr)$ that
absorb a chunk once it has reached its destination $d$ at some step $t<m_s$. For
$v_i\in\tilde V$ we write $\tilde\delta^+(v_i)$ and $\tilde\delta^-(v_i)$ for its
outgoing and incoming arcs.

\smallskip\noindent
\textit{Commodities and chunk budget ---}
In an \textsc{AllToAll} collective, each tensor $c{\in}C$ has source $src(c)$ and destinations $dst(c){=}V{\setminus}\{src(c)\}$. We route at the granularity of ordered GPU pairs, i.e., commodities
\begin{smallmath}
\setlength{\abovedisplayskip}{3pt}
\setlength{\belowdisplayskip}{3pt}
\mathcal K \;=\; \bigl\{\,k=(s,d)\;:\;s\in V,\ d\in dst(s)\,\bigr\},
\end{smallmath}%
where commodity $k=(s,d)$ delivers $K\in\mathbb Z_{>0}$ chunks from the source node $s_0$ to its super-sink $\sigma_d$. The commodities sharing a common
source $s=src(c)$ together form the flow of tensor $c$; this grouping is what the
decomposition of \Cref{app:a2a:dw} exploits.

\vspace{-3pt}
\subsection{Compact mixed-integer model}
\label{app:a2a:compact}

Let $f^{k}_{a_i}\in\{0,\dots,K\}$ be the integer number of chunks of
commodity $k\in\mathcal K$ routed on arc $a_i\in\tilde A$, and let
$\chi\ge 0$ be a global \emph{congestion} multiplier. The compact model minimises
congestion subject to delivering $K$ chunks per commodity, with every link load
held below $c_a\,\chi$:
\begin{smallalign}
&\min\quad \chi \span\span\span \label{cmp:obj}\\
&\textstyle\sum_{a_i\in\tilde\delta^-(v_i)} f^{k}_{a_i}
   \;-\;\sum_{a_i\in\tilde\delta^+(v_i)} f^{k}_{a_i}
   \;=\; b^{k}_{v_i}
   & & \forall k\in\mathcal K,\ \forall v_i\in\tilde V, \label{cmp:cons}\\
&\textstyle\sum_{k\in\mathcal K}\ \sum_{a_i\in\tilde A_a} f^{k}_{a_i}
   \;\le\; c_{a}\,\chi
   & & \forall a\in A, \label{cmp:cong}\\
& \textstyle f^{k}_{a_i}\in\{0,\dots,K\},\quad \chi\ge 0
   & & \forall k\in\mathcal K,\ \forall a_i\in\tilde A, \label{cmp:int}
\end{smallalign}
where the supply/demand vector encodes the per-commodity demand
$b^{k}_{v_i}\,=\,K(\mathbf{1}[v_i{=}\sigma_d]{-}\mathbf{1}[v_i{=}s_0])$.
Constraints~\eqref{cmp:cons} are per-commodity flow conservation on $\tilde G$,
and constraints~\eqref{cmp:cong} make the maximum normalised link load equal to
$\chi$ by summing every step-indexed copy of a link against its capacity $c_a$;
the step budget enters implicitly through the truncation of $\tilde A$ at $m_s$.
Since every commodity routes exactly $K$ chunks, the delivered throughput at
optimum is $K/\chi^\star$, where $\chi^\star$ is the optimal congestion. The
overall throughput is recovered by scanning the chunk budget
\begin{smallalign}
\textstyle z_{\max}^{\mathrm{MILP}} \;=\; \max_{K\in\mathbb Z_{>0}} \frac{K}{\chi^{\star}(K)}.
\label{cmp:scanK}
\end{smallalign}

\vspace{-3pt}
\subsection{Linear relaxation}
\label{app:a2a:relax}

Relaxing the integrality requirement~\eqref{cmp:int} to $f^{k}_{a_i}\ge 0$
makes the chunk budget $K$ redundant: the model is scale-invariant in $K$, so we
fix $K=1$ and let each commodity deliver a common rate $z=1/\chi$ to its sink.
Writing the relaxation directly on the flow-conservation constraints yields the
max--min concurrent-flow \gls{lp}, denoted $(\mathrm{CF\text{-}LP})$:
\begin{smallalign}
\setlength{\abovedisplayskip}{3pt}
\setlength{\belowdisplayskip}{3pt}
&\max\quad z \span\span\span \label{cf:obj}\\
&\textstyle \sum_{a_i\in\tilde\delta^-(v_i)} f^{(s,d)}_{a_i}
   \;-\;\sum_{a_i\in\tilde\delta^+(v_i)} f^{(s,d)}_{a_i}
   \;=\; b^{(s,d)}_{v_i}(z)
   & & \textstyle \forall (s,d)\in\mathcal K,\ \forall v_i\in\tilde V, \label{cf:cons}\\
&\textstyle \sum_{(s,d)\in\mathcal K}\ \sum_{a_i\in\tilde A_a} f^{(s,d)}_{a_i}
   \;\le\; c_a
   & & \textstyle \forall a\in A, \label{cf:cap}\\
& f^{(s,d)}_{a_i}\ \ge\ 0
   & & \textstyle \forall (s,d)\in\mathcal K,\ \forall a_i\in\tilde A, \label{cf:bound}
\end{smallalign}

where the (now rate-parameterised) demand vector is
$b^{(s,d)}_{s_0}=-z$, $b^{(s,d)}_{\sigma_d}=+z$, and $0$ otherwise.
Constraints~\eqref{cf:cons} are the per-commodity flow conservation, which tie
the rate routed to each sink to the common objective $z$;
constraints~\eqref{cf:cap} are the (coupling) link-capacity bounds; and
\eqref{cf:bound} is non-negativity. Maximising $z$ therefore maximises the
minimum per-commodity throughput.

\vspace{-3pt}
\subsection{Dantzig--Wolfe decomposition recovers the master}
\label{app:a2a:dw}
We apply Dantzig--Wolfe decomposition to $(\mathrm{CF\text{-}LP})$ tensor by
tensor, grouping the commodities that share a common source $s=src(c)$ into one
block per tensor $c$. The coupling link-capacity rows~\eqref{cf:cap} are kept in
the master, while the conservation rows~\eqref{cf:cons} define the blocks. Every
block flow is a non-negative combination of source-rooted trees
$T_c\in\mathcal T_c$: choosing one $src(c)\!\to\!d$ route for each destination
$d\in dst(c)$ yields one such tree, with $\beta_a^{T_c}$ counting the paths of
$T_c$ that load link $a$. Substituting the structure weights $x_{T_c}\ge 0$ for
the arc flows eliminates the conservation rows and reproduces, line for line,
the extensive master and its pricing subproblem; we therefore solve the
Restricted Master Problem and price columns exactly as described in
\Cref{subsec:a2a_optimization}.
\endgroup

\end{document}